\documentclass[graybox, UKenglish]{svmult}

\usepackage{fullpage}

\usepackage[justification=centering]{caption}
\usepackage{type1cm}        
\usepackage{makeidx}         
\usepackage{graphicx}        
\usepackage{multicol}        
\usepackage[bottom]{footmisc}

\usepackage{newtxtext}       %
\usepackage[varvw]{newtxmath}       

\makeindex             

\usepackage{csquotes}
\usepackage{mathtools}
\usepackage{stmaryrd}
\usepackage{tikz}
\usepackage{todonotes}
\usepackage{pifont}
\usepackage{hhline}
\usepackage{colortbl}
\usepackage{booktabs}
\usepackage{hyperref}

\usetikzlibrary{arrows, arrows.meta}
\usetikzlibrary{backgrounds}
\usetikzlibrary{calc}
\usetikzlibrary{decorations.markings, decorations.pathmorphing, decorations.pathreplacing}
\usetikzlibrary{fit}
\usetikzlibrary{matrix}
\usetikzlibrary{patterns}
\usetikzlibrary{positioning}
\usetikzlibrary{shapes}
\usetikzlibrary{tikzmark}

\usepackage{float}

\usepackage{bussproofs}

\newcommand\cutout[1]{}
\newcommand\eat[1]{}

\providecommand{\dotdiv}{
  \mathbin{
    \vphantom{+}
    \text{
      \mathsurround=0pt 
      \ooalign{
        \noalign{\kern-.35ex}
        \hidewidth$\smash{\cdot}$\hidewidth\cr 
        \noalign{\kern.35ex}
        $-$\cr 
      }%
    }%
  }%
}

\usepackage{mathtools} 
\usepackage{stmaryrd} 

\theoremstyle{claimstyle}

\newcommand{\ra}{\rightarrow}
\newcommand{\Iff}{\Leftrightarrow}
\renewcommand{\E}{\exists}
\newcommand{\A}{\forall}

\renewcommand{\phi}{\varphi}
\renewcommand{\theta}{\vartheta}
\renewcommand{\emptyset}{\varnothing}
\renewcommand{\epsilon}{\varepsilon}
\renewcommand{\AA}{{\mathfrak A}}
\newcommand{\BB}{{\mathfrak B}}
\newcommand{\N}{{\mathbb N}}

\newcommand{\FO}{{\rm FO}}

\newcommand{\posLFP}{{\rm posLFP}}
\DeclareMathOperator{\PL}{PL}

\DeclareMathOperator{\Lit}{\mathrm{Lit}}
\DeclareMathOperator{\nnf}{\mathrm{nnf}}

\newcommand{\entails}{\models}

\renewcommand{\bar}{\overline}

\newcommand{\Access}{\mathbb{A}}
\newcommand{\Trop}{\mathbb{T}}
\newcommand{\Vit}{\mathbb{V}}
\newcommand{\Nat}{\mathbb{N}}
\newcommand{\Ninf}{\mathbb{N}^\infty}
\newcommand{\Bool}{\mathbb{B}}

\newcommand{\Lukas}{\mathbb{L}}
\newcommand{\Doubt}{\mathbb{D}}
\newcommand{\Why}{\textrm{Why}}
\newcommand{\WW}{\mathbb{W}}

\newcommand{\PosBool}{\textrm{PosBool}}
\newcommand{\Sorb}{\mathbb{S}}
\newcommand{\Sinf}{\mathbb{S}^\infty}
\newcommand{\Ss}{\ensuremath{\mathcal{S}}}
\newcommand{\calS}{\mathcal{S}}
\newcommand{\Semiring}{\ensuremath{\calS}}
\newcommand{\Semi}{\Semiring}

\newcommand*{\ext}[1]{[\![ #1 ]\!]}

\newcommand{\Pub}{\mathsf{P}}
\newcommand{\Cnf}{\mathsf{C}}
\newcommand{\Sec}{\mathsf{S}}
\newcommand{\Tsec}{\mathsf{T}}

\newcommand{\Inf}{\bigsqcap}
\newcommand{\Sup}{\bigsqcup}
\newcommand{\meet}{\sqcap}
\newcommand{\join}{\sqcup}

\newcommand*{\Tt}{\ensuremath{\mathcal{T}}}

\newcommand*{\R}{\ensuremath{\mathbb{R}}}

\renewcommand{\phi}{\varphi}

\newcommand{\gothA}{\mathfrak{A}}
\newcommand{\gothE}{\mathfrak{E}}

\newcommand{\gothp}{\mathfrak{p}}
\newcommand{\gothq}{\mathfrak{q}}
\newcommand{\calF}{\mathcal{F}}

\newcommand{\calH}{\mathcal{H}}
\newcommand{\calM}{\mathcal{M}}
\newcommand{\calC}{\mathcal{C}}

\newcommand{\FactsA}{\mathsf{Facts}_A}
\newcommand{\LitA}{\mathsf{Lit}_A}
\newcommand{\NegFactsA}{\mathsf{NegFacts}_A}

\newcommand{\LitV}{\mathsf{Lit}_V}
\newcommand{\piA}{\pi_{_\gothA}}
\newcommand{\pisA}{\pi_{_{\#\gothA}}}
\newcommand{\Api}{\gothA_\pi}
\newcommand{\Abeta}{\gothA_\beta}
\newcommand{\Modpi}{\mathsf{Mod}_{\pi}}

\newcommand{\refi}[2]{#1|_{_#2}}
\newcommand{\domi}{\mathsf{dominant}}
\newcommand{\noVdom}{\mathsf{noVdom}}
\newcommand{\notdom}{\mathsf{notdom}}
\newcommand{\denydom}{\mathsf{denydom}}
\newcommand{\nn}[1]{\bar{#1}}

\newcommand{\nnp}{\nn{p}}
\newcommand{\nnq}{\nn{q}}
\newcommand{\nnr}{\nn{r}}
\newcommand{\nns}{\nn{s}}
\newcommand{\nnt}{\nn{t}}
\newcommand{\nnX}{\nn{X}}
\newcommand{\nnx}{\nn{x}}
\newcommand{\nnz}{\nn{z}}
\newcommand{\Fuzz}{\mathbb{F}}

\newcommand{\poly}[2]{#1\lbrack#2\rbrack}
\newcommand{\sem}[2]{\lbrack\!\lbrack#1\rbrack\!\rbrack_{#2}}

\newcommand{\ModDef}{model-defining}
\newcommand{\bfModDef}{\textbf{model-defining}}
\newcommand{\ModComp}{model-compatible}
\newcommand{\bfModComp}{\textbf{model-compatible}}

\newcommand{\val}[1]{}  
\newcommand{\erich}[1]{}

\begin{document}

\title*{Provenance Analysis and Semiring Semantics for First-Order Logic}
\author{Erich Gr{\"a}del\orcidID{0000-0002-8950-9991} and\\ Val Tannen\orcidID{0009-0008-6847-7274}}
\institute{Erich Gr{\"a}del \at RWTH Aachen University, Germany, \email{graedel@logic.rwth-aachen.de}
\and Val Tannen \at Univ.~of~Pennsylvania, U.S.A. \email{val@cis.upenn.edu}}
%
%
\maketitle

\abstract{
A provenance analysis for a query evaluation or a model checking computation extracts information on how its result depends on the atomic facts of the model or database. Traditional work on data provenance was, to a large extent, restricted to positive query languages or the negation-free fragment of first-order logic and showed how provenance abstractions can be usefully described as elements of commutative semirings --- most generally as multivariate polynomials with positive integer coefficients.
We describe and evaluate here a provenance approach for dealing with negation, based on quotient semirings of polynomials with dual
indeterminates. This not only provides a semiring provenance analysis for full first-order logic (and other logics  and query languages with negation) but also permits a reverse provenance analysis, i.e., finding models that satisfy various properties under given provenance tracking assumptions. We describe the potential for applications to explaining missing query answers or failures of integrity constraints,
and to using these explanations for computing repairs.
This approach also is the basis of a systematic study of semiring semantics in a broad logical context.
}

\section{Introduction}

\textbf{Semiring provenance} was originally developed for positive database query languages  by Green, Karvounarakis, and Tannen \cite{GreenKarTan07}.
It is based on the idea to annotate the atomic facts in a database by values in some commutative semiring, and to propagate these annotations through a 
query, keeping track whether information is used
alternatively (as in disjunctions or existential quantifications) or jointly (as in conjunctions or universal quantifications).
From this baseline, we have started in \cite{GraedelTan17} to investigate 
a new approach to the provenance analysis of model checking for languages with negation, and in particular full first-order logic (FO).
This approach is based on transformations to negation normal form, quotient semirings of polynomials with dual indeterminates, and a 
close relationship to semiring valuations of games \cite{GraedelTan20}. Since then, semiring provenance has been extended
to a systematic investigation of \emph{semiring semantics} for many logical systems, including first-order logic,
modal logic, description logics, guarded logic and fixed-point logic 
\cite{BourgauxOzaPenPre20, DannertGra19, DannertGra20,DannertGraNaaTan21}
and also to a general method for strategy analysis in games \cite{GraedelTan20,GraedelLucNaa21}. 
The present paper is a thoroughly revised and considerably expanded version of our paper \cite{GraedelTan17} from 2017 which has previously only appeared as a preprint in arXiv --- 
although it has been the basis of much of the subsequent work
on provenance analysis and semiring semantics. We shall also give a brief overview on the 
work that has been done since 2017, and discuss some questions for future research.

\medskip\noindent
\textbf{Data provenance} is extremely useful in many computational disciplines.
Suppose that a computational process is applied to a complex input consisting
of multiple items. Provenance analysis allows us to understand how these
different input items affect the output of the process. It can be used to
answer questions of the following type:
\begin{itemize}
\item Which ones of input items are actually used 
in the computation of the output? 
\item Can the same output be obtained from
different combinations of input items? 
\item In how many different ways
can the same output be computed? 
\end{itemize}

As a consequence, provenance can be further
applied to issues such as deciding how much to trust the output,
assuming that we may trust some input items more than others, deciding
what clearance level is required for accessing the output, assuming that
we know access restrictions for the input items, or, assuming that  one has to pay
for the input items, how to minimize the cost of obtaining the output.
More generally, \emph{reverse} provenance analysis
allows us to find input data (here first-order structures) that satisfy 
various properties under given provenance tracking assumptions.

It turns out that the questions listed above, as well as several other
questions of interest,
can be answered for database transformations (queries and views) via
interpretations in commutative semirings. In past work, the semiring
provenance approach has been applied to query and view languages such
as the positive relational algebra~\cite{GreenKarTan07,Green11},
nested relations/complex values 
(objects)~\cite{FosterGreTan08,Tannen13},
Datalog~\cite{GreenKarTan07,DeutchMilRoyTan14},
XQuery (for unordered XML)~\cite{FosterGreTan08},
relational algebra on $\mathbb{Z}$-annotated relations \cite{GreenIveTan11},
SQL aggregates~\cite{AmsterdamerDeuTan11b},
workflows with map-reduce modules~\cite{AmsterdamerDavDeuMilStoTan11},
and languages for data-centric (data-dependent) 
processes~\cite{DeutchMosTan15}. Moreover, the semiring approach has been 
successfully implemented in the software systems \textsf{Orchestra}
\cite{GreenKarIveTan07,IvesGreKarTayTanTalJacPer08,KarvounarakisIveTan10} and 
\textsf{Propolis}~\cite{DeutchMosTan15}. 
For a survey, see also \cite{Glavic21}.

However, for a long time, semiring provenance has essentially been restricted to negation-free query languages.
There have been algebraically interesting attempts to cover difference of relations
\cite{AmsterdamerDeuTan11,GeertsPog10,GeertsUngKarFunChr16,GreenIveTan11} but  
they had not resulted in systematic tracking of \emph{absent} or \emph{negative information}, and
for quite some time, this has been an obstacle for extending semiring provenance to other branches of logic
in computer science. Indeed, while there are many applications in databases where one can get quite far with
considering only positive information, logical applications in most other areas are based on
formalisms that use negation in an essential way.
A main objective of our new approach, proposed in
\cite{GraedelTan17}, was the extension of semiring provenance to more general logical formalisms, and in particular to full first-order logic.

\medskip\noindent\textbf{Provenance Semantics. }
We consider a non-standard semantics for first-order logic (FO)
that will help us to understand how a sentence $\phi$ ends up being true
in a finite structure~$\gothA$, i.e., 
whether $\gothA\models\phi$ holds or not (we call this  
\emph{provenance in model checking}).
The non-standard
semantics that we champion involves various \emph{commutative
  semirings}.  Here we strive to justify this choice.

First of all, the standard semantics for first-order logic maps formulae
to truth values in $\Bool=\{\bot,\top\}$, which 
form a commutative semiring with respect 
to the operations of disjunction and conjunction, with units $\bot$
and $\top$.
Second, in a provenance semantics we want to understand the
connections between the facts (positive or negative) that are embodied
in a model $\gothA$ and their use in a justification that
$\gothA\models\phi$. We can think of such a
justification as a disjunction-conjunction
\emph{proof tree} (an example appears in Sect.~\ref{subsec:ex}) or, equivalently, as 
a winning strategy in the model checking game for $\AA$ and $\phi$.   
If we had a provenance semantics
for model checking, it would, in particular, help us to count such proof trees or evaluation strategies.
This particular case suffices to suggest the semiring structure as well
as some ways in which such non-standard semantics can be quite different
from the standard one. 

Notice that a semiring semantics refines the classical Boolean semantics,
and formulae that are classically equivalent may become 
non-equivalent under a semantics that counts proof trees.
Indeed, already a sentence $\phi\vee\phi$ has in general
more proof trees than $\phi$.
We further illustrate this with the failure
of some of the usual logical equivalences
invoked in transforming sentences to \emph{prenex form}.

Let $\rho:=(\forall x\, \phi) \wedge \psi$ and $\sigma:=
\forall x\, (\phi \wedge \psi)$.  Every proof tree of $\rho$ can be
transformed into a proof tree of $\sigma$ by making copies of the
subtree rooted at $\psi$.  However, when $\psi$ has two or more
distinct proof trees we see that $\sigma$ can have strictly more proof
trees than $\rho$. Similarly we can argue that
$\forall x\, (\phi \vee \psi)$ can have strictly more proof trees
than $(\forall x\, \phi) \vee \psi$.
Now consider $\rho:=(\exists x\, \phi) \vee \psi$
and $\sigma:=\exists x\, (\phi \vee \psi)$. Let us write
$\phi(x)$ to show occurrences of $x$ in $\phi$. For simplicity
suppose that the model has exactly two elements, $a$ and $b$, and that
each of $\phi(a)$, $\phi(b)$, and $\psi$ has exactly one proof tree.
Then, $\rho$ will have three proof trees but $\sigma$ will have four.
Finally, we note that $(\exists x\, \phi) \wedge \psi$
and $\exists x\, (\phi \wedge \psi)$ have exactly the same number
of proof trees and this reflects the fact that multiplication distributes 
over addition. 

For other sentences, we can see that the number-of-proof-trees
constitutes a non-standard semantics for first-order sentences constructed
using disjunction, conjunction, existentials and universals,
because, moreover, addition and multiplication are
associative and commutative.

This discussion provides some partial justification for considering
commutative semirings as semantic domains. The rest of the justification
will follow from the subsequent development.

\section{Commutative semirings}\label{sec:commutative-semirings}
	
\begin{definition}[Semiring]
	A commutative semiring is an algebraic structure $\Semi = (S, +, \cdot, 0, 1)$ with $0  \neq 1$, such that $(S, +, 0)$ and $(S, \cdot, 1)$ are commutative monoids, $\cdot$ distributes over $+$, and $0 \cdot s = s \cdot 0 = 0$.
\end{definition}

A commutative semiring is  \emph{naturally ordered} (by addition) if $s \leq t :\Leftrightarrow \exists r (s+r = t)$ defines a partial order. 
Notice that $\leq$ is always reflexive and transitive, so a semiring is naturally ordered if, and only if, $\leq$ is antisymmetric, i.e.
$r\leq s$ and $s\leq r$ only hold for $s=r$. In particular, this excludes rings.
In this paper, we only consider commutative and naturally ordered semirings and simply refer to them as \emph{semirings}.  
A semiring $\Semi$ is \emph{idempotent} if $s+s=s$ for each $s \in S$ and \emph{multiplicatively idempotent} if $s \cdot s = s$ for all $s \in S$.
If both properties are satisfied, we say that $\Semi$ is fully idempotent.
Finally, $\Semi$ is \emph{absorptive} if $s + s t = s$ for all $s, t \in S$ or, equivalently, if multiplication is decreasing in $\Semi$, i.e. $st \leq s$ for $s, t \in S$. 
Every absorptive semiring is idempotent, and every idempotent semiring is naturally ordered.

\subparagraph*{Application semirings} There are many applications which can be modelled by semirings and provide useful practical information about the evaluation of a formula. 
\begin{itemize}
\item
The Boolean semiring $\Bool=(\Bool,\vee,\wedge,\bot,\top)$ is the standard habitat of 
logical truth.
\item A totally ordered set $(S, \leq)$ with least element $s$ and greatest element $t$ induces the \emph{min-max semiring} $(S, \max, \min, s, t)$. 
Specific important examples are the Boolean semiring, the fuzzy semiring $\Fuzz=([0,1],\max,\min,0,1)$, and the  
\emph{access control semiring}, also called the \emph{security semiring} \cite{FosterGreTan08}, which
is a min-max semiring with elements $0 < \Tsec < \Sec < \Cnf < \Pub=1$ where 0 stands for ``inaccessible'' (or ``false''),
$\Tsec$ is ``top secret'', $\Sec$ is ``secret'', $\Cnf$ is ``confidential'', and
$\Pub$ is ``public''. It is used for reasoning about access restrictions to atomic facts, and
the clearance levels that are necessary to verify the truth of a sentence under such restrictions.
\item A more general class (than min-max semirings) is the class of  \emph{lattice semirings} $(S,\meet,\join,s,t)$ induced by
a bounded distributive lattice $(S,\leq)$. Clearly, lattice semirings are fully idempotent.
\item The \emph{tropical semiring} $\Trop = (\R^\infty_+ , \min, +, \infty, 0)$ is used to annotate atomic facts with a cost for accessing them and 
to compute minimal costs for verifying a logical statement. It is not fully idempotent but absorptive.
\item The \emph{Viterbi semiring} $\Vit = ([0,1]_\R,\max,\cdot,0,1)$, which is in fact  isomorphic to $\Trop$ via $y \mapsto -\ln y$
can be used for  reasoning about confidence. 
\item An alternative semiring for reasoning about confidence scores is the \emph{Łukasiewicz semiring} $\Lukas= ([0,1]_\R, \max, \odot, 0, 1)$, where multiplication is given by $s \odot t = \max(s+t -1, 0)$. It is isomorphic to the semiring of doubt $\Doubt = ([0,1]_\R, \min, \oplus, 1, 0)$ with $s \oplus t = \min(s+t, 1)$. Also $\Lukas$ and $\Doubt$ are absorptive semirings. 
\item The \emph{natural semiring} $\N = (\N, +, \cdot, 0, 1)$ is used to count the number of proof trees or evaluation strategies that estabish the truth of a sentence. It is also important for bag semantics in databases. 
\end{itemize}
	
\subparagraph*{Provenance semirings}
Provenance semirings of polynomials provide information on which combinations of literals imply the truth of a formula.
The universal provenance semiring is the semiring $\N[X]$ of multivariate polynomials with indeterminates from $X$ and coefficients from $\N$. Other provenance 
semirings are obtained, for example, as quotient semirings of $\N[X]$ induced by congruences for idempotence and absorption. The resulting provenance values are less informative but their computation is more efficient. 
\begin{itemize}
	\item By dropping coefficients from $\N[X]$, we get the free idempotent semiring $\Bool[X]$ whose elements are (in one-to-one correspondence with) finite sets of monomials with coefficient 1. It is the quotient induced by $x + x \thicksim x$.
	\item If, in addition, exponents are dropped, we obtain the Why-semiring $\WW(X)$ of finite sums of monomials with coefficient 1 that are linear in each indeterminate.
	\item The free absorptive semiring $\Sorb(X)$ consists of $0,1$ and all antichains of monomials with respect to the absorption order $\succcurlyeq$. A monomial $m_1$ absorbs $m_2$, denoted $m_1 \succcurlyeq m_2$, if it has smaller exponents, i.e. $m_2 =m \cdot m_1$ for some monomial $m$.
	\item Finally, $(\PosBool(X),\vee,\wedge,\bot,\top)$ is the semiring
whose elements are classes of equivalent (in the usual sense) positive 
boolean expressions with boolean variables from $X$. Its elements
are in bijection with the positive boolean expressions in irredundant disjunctive normal form.
This is the lattice semiring freely generated by the set $X$.
It arises from $\Sorb(X)$ by dropping exponents.
\end{itemize}

\section{First-Order Logic Interpreted in Commutative Semirings}

We are interested in the provenance analysis of the model checking
computation of first-order sentences. Such a computation is nicely and
\emph{declaratively} driven by the structure of the sentence, and
thus amounts to a non-standard semantics for FO.  In its simplest
form, model checking takes as input a finite structure and
the input items are the various facts (positive or negative) which
hold in the model. We have found however that it pays to take a more
general approach and specify not a structure but just its (finite)
universe. This way we can track the use of positive and negative facts
in checking a sentence under multiple possible models on that universe.
This allows a certain amount of \emph{reverse analysis}: finding models
that satisfy useful constraints.

\subsection{Semiring interpretations}
Consider a finite relational vocabulary $\tau=\{R,S,\ldots\}$.
From this vocabulary and a finite, non-empty universe $A$ of \emph{ground
values} we construct the set $\FactsA(\tau)$ of all ground relational
atoms (facts) $R\bar a$, the set $\NegFactsA(\tau)$ of all negated
facts $\neg R\bar a$ and thus the set
$\LitA(\tau)=\FactsA(\tau)\cup\NegFactsA(\tau)$ of all \emph{literals},
positive and negative facts, over $\tau$ and $A$.
By convention we will identify $\neg\neg R\bar a$ with  $R\bar a$
so the negation of a literal is again a literal.

Any finite  structure $\gothA = (A,R^{\gothA},S^{\gothA},\ldots)$ 
with universe $A$ makes some of these
literals true and the remaining ones false. Note, however, that
much of the development does not assume a specific model, and this 
can be usefully exploited.
Let $\Semi=(S,+,\cdot,0,1)$ be a commutative semiring. Very roughly
speaking, $0\in S$ is intended to interpret false assertions,
while an element $s\neq0$ in $\Semi$ provides a ``nuanced'' or ``annotated" interpretation
for true assertions.

Next, $\Semi$-interpretations will map literals to elements of $\Semi$
and are then extended to all formulae.
Disjunction and existential quantification are interpreted by the
addition operation of $\Semi$.
Conjunction and universal quantification are interpreted by the
multiplication operation of $\Semi$. 
For quantifiers, the finiteness of the universe
$A$ of ground values will be essential. Extensions to infinite universes are possible
for semirings with appropriate infinitary addition and multiplication operations,
see \cite{BrinkeGraMrkNaa24}, but they will not be considered in this paper.
For negation we use the well-known syntactic transformation to 
\emph{negation normal form (NNF)}, denoted $\psi\mapsto\nnf(\psi)$.
Note that $\nnf(\psi)$ is a formula constructed from 
literals (positive and negative facts) and equality/inequality atoms 
using just $\wedge,\vee,\exists,\forall$. 

\begin{definition}
\label{def:sem}
An \textbf{$\Semi$-interpretation} is a mapping $\pi:\LitA(\tau)\rightarrow S$.
This extends to valuations $\pi \ext{\phi(\bar a)}$
of any instantiation of a formula $\phi(\bar x) \in \FO(\tau)$, 
by a tuple $\bar a \subseteq A$.
We first extend $\pi$ by mapping equalities and inequalities to their truth values, by 
setting $\pi \ext{a = a}:=  1$ and $\pi\ext{a=b} := 0$ for $a\neq b$ (and analogously for
inequalities). Further
\begin{alignat*}{3}
\pi \ext{\psi \lor \phi} &:= \pi \ext{\psi} + \pi \ext{\phi} &\quad\quad\quad \pi \ext{\psi \land \phi} &:= \pi \ext{\psi} \cdot \pi \ext{\phi} \\
\pi \ext{\exists x \, \psi(x)} &:= \sum\nolimits_{a \in A} \pi \ext{\psi(a)} &\quad\quad\quad \pi \ext{\forall x \, \psi(x)} &:= \prod\nolimits_{a \in A} \pi \ext{\psi(a)}\\
\pi\ext{\neg\phi}&:= \pi\ext{\nnf(\neg\phi)}&&
\end{alignat*}
\end{definition}

Notice that equality and inequality atoms are interpreted in $\Semi$ as $0$ or $1$, i.e.,
their provenance is not tracked. One could give a similar treatment to other
such relations with ``fixed'' meaning, e.g., assuming an ordering on $A$, 
however, we omit this here.
By a trivial induction it follows that, as intended, it suffices to consider formulae 
in negation normal form. 

\begin{proposition} 
\label{prop:NNF}
\hspace*{5mm}
$\pi\ext{\phi} ~=~ \pi\ext{\nnf(\phi)}$.
\end{proposition}

A useful consequence of Proposition~\ref{prop:NNF} is that we can prove
further results by induction on formulas in NNF, and hence avoid the
non-atomic negation altogether.

\begin{proposition}[Fundamental Property]
\label{prop:hom}
Let $h:\Semi_1\rightarrow \Semi_2$ be a semiring homomorphism
and let $\pi_1:\LitA(\tau)\rightarrow \Semi_1$ and
$\pi_2:\LitA(\tau)\rightarrow \Semi_2$ be interpretations such that
$h\circ\pi_1=\pi_2$. Then, for every sentence $\phi\in\FO(\tau)$ we have
$h(\pi_1\sem{\phi}{}) ~=~ \pi_2\sem{\phi}{}$. As diagrams

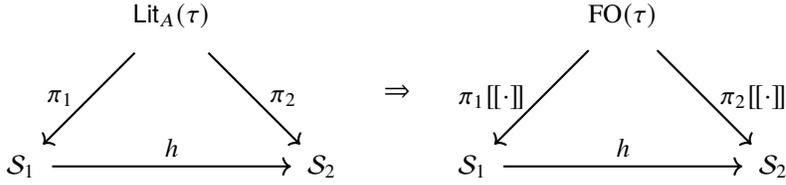
\begin{tikzpicture}
\begin{scope}
  \node[shape=circle] (A) at (0,0) {$\Semi_1$};
  \node[shape=circle] (B) at (2,2) {$\LitA(\tau)$};
  \node[shape=circle] (C) at (4,0) {$\Semi_2$};

 \draw[->,thick] 
      (B) edge node[left] {$\pi_1\:$} (A);
 \draw[->,thick] 
      (B) edge node[right] {$\:\pi_2$} (C);
 \draw[->,thick] (A) edge node[above] {$h$} (C);

\node[shape=circle] (D) at (5,1) {$\Rightarrow$};

  \node[shape=circle] (A) at (6,0) {$\Semi_1$};
  \node[shape=circle] (B) at (8,2) {$\FO(\tau)$};
  \node[shape=circle] (C) at (10,0) {$\Semi_2$};

 \draw[->,thick] 
      (B) edge node[left] {$\pi_1\ext{\cdot}\:$} (A);
 \draw[->,thick] 
      (B) edge node[right] {$\:\pi_2\ext{\cdot}$} (C);
 \draw[->,thick] (A) edge node[above] {$h$} (C);
 \end{scope}
\end{tikzpicture}
\end{proposition}

\begin{proof}
By Proposition~\ref{prop:NNF} the proof can proceed by induction
on formulae in NNF. For example
$h(\pi_1\sem{\phi\wedge\psi}{}) =
h(\pi_1\sem{\phi}{} \cdot_{_1} \pi_1\sem{\psi}{}) = 
h(\pi_1\sem{\phi}{}) \cdot_{_2} h(\pi_1\sem{\psi}{}) = 
\pi_2\sem{\phi}{} \cdot_{_2} \pi_2\sem{\psi}{} = 
\pi_2\sem{\phi\wedge\psi}{}$.
\end{proof}

The somewhat bombastic name ``fundamental property'' is motivated by
two observations. First, the property checks that the
definition of our semantics is nicely compositional. Second,
the property plays a central role in a strategy that we have widely
applied with query languages in databases: compute provenance
as generally as (computationally) feasible, then specialize via
homomorphisms to
coarser-grained provenance, or to specific domains, e.g., count, trust, cost
or access control.

\subsection{Intermezzo: Positive Semirings}
We say that a semiring $\Semi$ has \emph{divisors of 0} if there exist non-zero
elements $s,t\in S$ such that $st=0$. Among
the semirings described in Sect.~\ref{sec:commutative-semirings}, only the
Lukasiewicz semiring $\Lukas$ and its isomorphic variant $\Doubt$ have
divisors of 0. Indeed in $\Lukas$, we have that $s \odot t = 0$ if, and only if $s+t\leq 1$.
We shall discuss in Sect.~\ref{sec:prov-semiring-FOL}
further interesting semirings with divisors of 0.

A semiring $\Semi$ is \emph{$+$-positive} if $s+t=0$ implies $s=0$ and
$t=0$.  All semirings described 
in Sect.~\ref{sec:commutative-semirings} are +-positive, but rings
are not. Finally, a semiring is 
\emph{positive} if it is $+$-positive and has no
divisors of $0$.

\begin{proposition}
\label{prop:positive-semiring}
A semiring $\Semi$ is positive if, and only if,
$~~\dagger_{_\Semi}:S\rightarrow\Bool~~$ defined by

\hspace*{1cm}
\begin{minipage}{0.5\textwidth}
$$
\dagger_{_\Semi}(s)= \begin{cases}
              \top & \mathrm{if~} s\neq0\\
              \bot & \mathrm{if~} s=0
              \end{cases}
$$
\end{minipage}
\begin{minipage}{0.4\textwidth}
is a homomorphism.
\end{minipage}
\end{proposition}

\subsection{Sanity Checks}

Let~~$\gothA = (A,R^{\gothA},S^{\gothA},\ldots)$~~be a 
(finite) $\tau$-model. 
The \textbf{canonical truth interpretation} for $\gothA$
is, of course, ~$\piA:\LitA(\tau)\rightarrow\Bool$ where
$$
\piA(L) ~=~ \begin{cases}
            \top & \mathrm{if~} \gothA\models L\\
            \bot & \mathrm{otherwise}
            \end{cases}
$$

Earlier we have discussed the ``number of proof trees''
as a non-standard semantics for \FO-model checking.
This is also captured by interpretations in a semiring.
The \textbf{canonical counting interpretation} for $\gothA$
is $\pisA : \LitA\rightarrow\Nat$ where
$$
\pisA(L) ~=~ \begin{cases}
              1 & \mathrm{if~} \gothA\models L\\
              0 & \mathrm{otherwise}
              \end{cases}
$$

\begin{proposition}[sanity checks]
\label{prop:sanity}
For any first-order sentence $\phi$ we have
$\gothA\models\phi$ if, and only if,
$\piA\sem{\phi}{}=\top$.
Moreover, $\pisA\sem{\phi}{}$ is the number of proof trees that
witness $\gothA\models\varphi$.
\end{proposition}

Now, let $\Semi$ be a commutative semiring, and let $\pi:\LitA(\tau)\rightarrow S$
be a $\Semi$-interpretation.  As we have indicated, for a sentence
$\phi$ we intend to interpret $\pi\sem{\phi}{}=0$ as ``$\phi$ is false
for $\pi$'', while $\pi\sem{\phi}{}=s\neq0$ is interpreted as ``$\pi$ makes $\phi$
true with annotation $s$''. We examine
how this meshes with standard logical truth in a model.

\begin{definition}
\label{def:int-model-defining}
A $\Semi$-interpretation $\pi:\LitA(\tau)\rightarrow S$ is 
\bfModDef\
when, for each fact $R\bar a$, precisely one of the values $\pi(R\bar a)$ and $\pi(\neg R\bar a)$
is $0$.
Indeed, every \ModDef\ interpretation $\pi$ uniquely defines a 
model $\Api$ with universe $A$ such that for any literal
$L$ we have $\Api\models L$ if, and only if, $\pi(L)\neq0$.
\end{definition}

Both $\piA$ and $\pisA$ shown above are \ModDef\ and the model they
define is $\gothA$. If $\Semi$ is not $\Bool$ then several
\ModDef\ interpretations may define the same model. It is also
clear that any finite model can be defined by such an interpretation,
for any $\Semi$.

\begin{proposition}[another sanity check]
\label{prop:strong}
Let $\Semi$ be \emph{positive}, 
and let $\pi$ be a \ModDef\ $\Semi$-interpretation.
Then for any \FO-sentence $\phi$
$$
\Api\models\phi ~~\Leftrightarrow~~\pi\sem{\phi}{}\neq0
$$ 
\end{proposition}

\begin{proof}
By Proposition~\ref{prop:positive-semiring},
since $\Semi$ is positive, $\dagger_{_\Semi}$ is a homomorphism.
Since $\pi$ is \ModDef\ let $\gothA$ be the model defined by $\pi$.
Clearly, $\dagger_{_\Semi}\circ\pi$ is the canonical truth interpretation $\piA$. 
Applying Proposition~\ref{prop:hom}
we get $\dagger_{_\Semi}(\pi\sem{\phi}{}) = \piA\sem{\phi}{}$. 
Now the result follows from Proposition~\ref{prop:sanity}.
\end{proof}

In fact, we can refine the previous proposition as follows.

\begin{proposition}[refinement of Proposition~\ref{prop:strong}]\label{prop:refi}
For any semiring $\Semi$ (positive or not!),
for any \ModDef\ $\Semi$-interpretation $\pi$, and 
for any \FO-sentence $\phi$ we have
$$
\pi\sem{\phi}{}\neq0 ~~\Rightarrow~~ \Api\models\phi. 
$$ 
Moreover, a semiring $\Semi$ is positive if, and only if,
for any \ModDef\ $\Semi$-interpretation $\pi$
and any \FO-sentence $\phi$ we have
$$
\Api\models\phi ~~\Rightarrow~~\pi\sem{\phi}{}\neq0.
$$ 
\end{proposition}
\begin{proof} 
The first part of the proposition is a simple induction on $\phi$.
For the second 
implication we first prove that $\Semi$ has no divisors of $0$.
Suppose that $s,t\in \Semi$ are such that $s\neq0$, $t\neq0$ but
$st=0$. Consider $A=\{a_1,a_2\}$ and the \ModDef\
interpretation defined by $\pi(\neg Ra_1)=\pi(\neg Ra_2)=0$,
$\pi(Ra_1)=s$, $\pi(Ra_2)=t$ as well as the sentence
$\phi=Ra_1\wedge Ra_2$. We have $\Api\models\phi$ 
hence $\pi\sem{\phi}{}\neq0$, contradiction.

Next we prove that $\Semi$ is $+$-positive. Let $s,t\in \Semi$ be such 
that $s\neq0$ or $t\neq0$. Consider the same interpretation
$\pi$ as above, with the sentence 
$\psi=Ra_1\vee Ra_2$. We have $\Api\models\psi$ hence 
$0\neq\pi\sem{\psi}{}=s+t$.
\end{proof}

\subsection{``Consistency'' and ``completeness'' for semiring interpretations}

In the study of provenance we shall also have occasion to consider
interpretations that do not correspond to a single specific model
(as formalized in Definition~\ref{def:int-model-defining}). 
Additional issues arise for such interpretations.

An interpretation in which both $\pi\sem{\phi}{}\neq0$ and
$\pi\sem{\neg\phi}{}\neq0$ for some sentence $\phi$ is seemingly
``inconsistent''. On the other hand, an interpretation
in which both $\pi\sem{\phi}{}=0$ and
$\pi\sem{\neg\phi}{}=0$ for some sentence $\phi$ seems to to be
``incomplete''. 
\erich{I would omit this footnote}
Of course, neither of these situations arises for a \ModDef\
$\Semi$-interpretation when $\Semi$ is positive (by Proposition~\ref{prop:strong}).
We analyze each of these issues in turn for general
interpretations.

\begin{proposition}
\label{prop:K-consistency-one}
Let $\pi:\LitA(\tau)\rightarrow \Semi$ be a $\Semi$-interpretation.
If for every $L\in\LitA(\tau)$ at least one of $\pi(L)$
and $\pi(\neg L)$ is $0$
then there exists no sentence $\;\psi\;$ for which both 
$\pi\sem{\psi}{}\neq0$ and $\pi\sem{\neg\psi}{}\neq0$.
\end{proposition}

\begin{proof} Suppose that both $\pi\ext\psi$ and $\pi\ext{\neg\psi}$ are non-zero.
Since we assume $A$ to be finite, there exist finitely many sentences 
$\phi_1,\dots,\phi_k$ such that one of the values
$\pi\ext\psi$ and $\pi\ext{\neg\psi}$ is the sum of $\pi\ext{\phi_1},\dots,\pi\ext{\phi_k}$,
and the other is the product of $\pi\ext{\neg\phi_1},\dots\pi\ext{\neg\phi_k}$.
It follows that all values $\pi\ext{\neg\phi_i}$ are non-zero.
But by induction hypothesis, this implies that all values $\pi\ext{\phi_i}$, 
and hence also their sum, must be 0, so we have a contradiction.
\end{proof}

Observe that if at least one of $\pi\sem{\phi}{}$ or
$\pi\sem{\neg\phi}{}$ is $0$ then 
$\pi\sem{\phi}{}\cdot\pi\sem{\neg\phi}{}=0$. If $\Semi$ has no
divisors of 0 the converse holds as well. Although most of the examples
described in~Sect.~\ref{sec:commutative-semirings} 
are positive semirings, we are about to introduce,
in~Sect.~\ref{subsec:dual-ind-poly}, a semiring for \FO-provenance that
\emph{does} have divisors of $0$. For this reason we note also the
following:

\begin{proposition}
\label{prop:K-consistency-two}
Let $\pi:\LitA(\tau)\rightarrow \Semi$ be an $\Semi$-interpretation.
If for every $L\in\LitA$ we have $\pi(L)\cdot\pi(\neg L)=0$
then for any sentence $\psi$ we have
$\pi\sem{\psi}{}\cdot\pi\sem{\neg\psi}{}=0$.
\end{proposition} 

\begin{proof} If $\psi$ is not a literal, then there exists sentences $\phi_1,\dots,\phi_k$
such that one of the values
$\pi\ext\psi$ and $\pi\ext{\neg\psi}$ is the sum of $\pi\ext{\phi_1},\dots,\pi\ext{\phi_k}$,
and the other is the product of $\pi\ext{\neg\phi_1},\dots\pi\ext{\neg\phi_k}$.
By induction hypothesis we have that $\pi\ext{\phi_i}\cdot\pi\ext{\neg\phi_i}=0$ for all $i\leq k$.
It follows that
\begin{align*} \pi\ext{\psi}\cdot\pi\ext{\neg\psi}= 
& \sum_{i\leq k}\pi\ext{\phi_i}\cdot\prod_{j\leq k}\pi\ext{\neg\phi_j} =
\sum_{i\leq k} \Bigl(\pi\ext{\phi_i}\cdot\prod_{j\leq k}\pi\ext{\neg\phi_j}\Bigr)\\
&=\sum_{i\leq k} \Bigl(\pi\ext{\phi_i}\cdot\pi\ext{\neg\phi_i}\cdot\prod_{j\neq i}\pi\ext{\neg\phi_j}\Bigr)
=0.
\end{align*}
\end{proof}

Proposition~\ref{prop:K-consistency-one} and~Proposition~\ref{prop:K-consistency-two}
hold in arbitrary semirings and each supports a kind of ``consistency'', with
the two kinds coinciding when the semiring has no divisors of $0$.

Turning to ``completeness'', note that if both $\pi\sem{\phi}{}$ and
$\pi\sem{\neg\phi}{}$ are $0$ then
$\pi\sem{\phi}{}+\pi\sem{\neg\phi}{}=0$.  If $\Semi$ is +-positive then
the converse holds as well.  However, for arbitrary $\Semi$,
neither an analog of
Proposition~\ref{prop:K-consistency-one} nor one of
Proposition~\ref{prop:K-consistency-two} holds. Indeed, let
$\Semi=\mathbb{Z}_4$. Consider the vocabulary consisting of one unary
relation symbol $R$ and let $A=\{a_1,a_2\}$.  For the interpretation
given by $\pi(\neg Ra_1)=\pi(\neg Ra_2)=\pi(Ra_1)=\pi(Ra_2)=2$
and the sentence $\phi=Ra_1\wedge Ra_2$ we have
$\pi\sem{\phi}{}=\pi\sem{\neg\phi}{}=0$.

Instead, we have the following for positive semirings.

\begin{proposition}
\label{prop:K-completeness}
Let $\pi:\LitA(\tau)\rightarrow \Semi$ be an $\Semi$-interpretation
into a positive semiring.
If for every $L\in\LitA(\tau)$ we have $\pi(L)\neq0$ or $\pi(\neg L)\neq0$
(equivalently, $\pi(L)+\pi(\neg L)\neq0$)
then for any sentence $\psi$ we have
$\pi\sem{\psi}{}\neq0$ or $\pi\sem{\neg\psi}{}\neq0$
(equivalently, $\pi\sem{\psi}{}+\pi\sem{\neg\psi}{}\neq0$).
\end{proposition}

\begin{proof} Towards a contradiction, suppose that $\pi\ext\psi=\pi\ext{\neg\psi}=0$.,
As in the two previous proofs, take
$\phi_1,\dots,\phi_k$ such that one of the values
$\pi\ext\psi$ and $\pi\ext{\neg\psi}$ is the sum of $\pi\ext{\phi_1},\dots,\pi\ext{\phi_k}$,
and the other is the product of $\pi\ext{\neg\phi_1},\dots\pi\ext{\neg\phi_k}$.
Since $\Semi$ has no divisors of 0, it follows that 
$\pi\ext{\neg\phi_i}=0$ for at least one $i\leq k$. 
By induction hypothesis, $\pi\ext{\phi_i}\neq 0$, which, by +-positivity,
contradicts the assumption that $\pi\ext\psi= 0$.
\end{proof}

\subsection{Proof trees}\label{sec:prooftrees}

Reasoning about the proof trees that a particular semiring interpretation
admits for a given first-order sentence is an important aspect of provenance analysis. We shall prove that the provenance value of
every sentence is the same as the sum of the valuations of its proof trees.
To establish this result, we have to provide a precise defintion
of proof trees and their valuations.

\medskip An \emph{evaluation tree} for a sentence $\psi\in\FO$ and
a semiring interpretation $\pi:\LitA(\tau)\ra\Ss$ (into an arbitrary semiring $\Ss$) is a directed tree $\Tt$, whose nodes are (labelled by) occcurrences\footnote{Notice that we consider different 
occurrences of the same subformula as separate objects. In particular, a sentence $\phi\lor\phi$
has twice as many evaluation trees as $\phi$.}  of formulae $\phi(\bar a)$, where $\phi(\bar x)$ is a subformula of $\psi$ whose free variables $\bar x$
are instantiated by a tuple $\bar a$ of elements from $A$, such that the following conditions hold.

\begin{itemize}
\item The root of $\Tt$ is  $\psi$.
\item A node $\phi\lor\theta$ has one child which is either $\phi$ or $\theta$.
\item A node $\phi\land\theta$ has two children $\phi$ and $\theta$.
\item A node $\E y\phi(\bar a,y)$ has one child $\phi(\bar a,b)$
for some $b\in A$.
\item A node $\A y\phi(\bar a,y)$ has the children $\phi(\bar a,b)$
for all $b\in A$.
\item The leaves of $\Tt$ are literals $L\in\LitA(\tau)$.
\end{itemize}

For any literal $L$, let 
$\#_L(\Tt)$ be the number of occurrences of
$L$ in $\Tt$. The valuation $\Tt$ is
\[ \pi(\Tt):=\sum_{L\in\LitA(\tau)} \pi(L)^{\#_L(\Tt)}.\]

A \emph{proof tree} for $\pi$ and  $\psi\in\FO$ is
an evaluation tree $\Tt$ with $\pi(\Tt)\neq 0$.
If $\pi$ is clear from the context, we write $T(\psi)$ for the set of all proof trees for $\pi$ and $\psi$.

\begin{theorem}\label{thm:prooftrees}
For every semiring interpretation 
$\pi:\LitA(\tau)\rightarrow \Ss$ and every sentence
$\psi\in \FO(\tau)$, we have that
\[ \pi\sem{\psi}{} =\sum_{\Tt\in T(\psi)} \pi(\Tt).\]
\end{theorem}

\begin{proof}
We proceed by induction on $\psi$. 
\begin{itemize}
\item Let $\psi$ be a literal. If $\pi(\psi)=0$ then $\psi$ has no proof tree, so the sum over the valuations of its proof trees is 0.
Otherwise $\psi$ has precisely one proof tree which is the literal itself. In both cases, the desired equality holds trivially.
\item Let $\psi=\phi\lor\theta$. A proof tree $\Tt$ for $\psi$ has the root $\psi$ followed by a proof tree $\Tt'$ for either $\phi$ or for $\theta$; clearly $\pi(\Tt)=\pi(\Tt')$.
Thus 
\[ \pi\ext\psi=\pi\ext{\phi}+\pi\ext{\theta}=
\sum_{\Tt'\in T(\phi)} \pi(\Tt') +\sum_{\Tt'\in T(\theta)} \pi(\Tt') =\sum_{\Tt\in T(\psi)} \pi(\Tt). \]
\item Let $\psi=\phi\land\theta$. A proof tree $\Tt$ for $\psi$ has the root $\psi$, attached to which are a proof tree $\Tt'$ for $\phi$ and a proof tree $\Tt''$ for $\theta$. We can thus identify every $\Tt\in T(\psi)$ with a pair $(\Tt',\Tt'')\in T(\phi)\times T(\theta)$,
and since $\#_L(\Tt)=\#_L(\Tt')+\#_L(\Tt'')$ for every literal $L$ we have that, $\pi(\Tt)=\pi(\Tt')\pi(\Tt'')$. It follows that
\begin{align*} \pi\ext\psi&=\pi\ext{\phi}\cdot\pi\ext{\theta}=
\sum_{\Tt'\in T(\phi)} \pi(\Tt') \cdot \sum_{\Tt''\in T(\theta)} \pi(\Tt'')\\ 
&=  \sum_{(\Tt',\Tt'')\in T(\psi)} \pi(\Tt')\pi(\Tt'')
= \sum_{\Tt\in T(\psi)} \pi(\Tt). 
\end{align*}
\item  If $\psi=\E y\phi(y)$, then a proof tree $\Tt$ for $\psi$ consists of the root $\psi$, attached to which is a proof tree $\Tt_a$ for $\phi(a)$, for some $a\in A$. Clearly $\pi(\Tt)=\pi(\Tt_a)$.
It follows that
\[ \pi\ext\psi=\sum_{a\in A}\pi\ext{\phi(a)}=
\sum_{a\in A}\ \ \sum_{\Tt_a\in T(\phi(a))} \pi(\Tt_a)=
\sum_{\Tt\in T(\psi)} \pi(\Tt). \]
\item Let finally $\psi=\A y\phi(y)$. A proof tree $\Tt$ for $\psi$ consists of the root $\psi$ attached to which are proof trees $\Tt_a$ for $\phi(a)$, for all $a\in A$, so we can identify every proof tree $\Tt\in T(\psi)$ with the tuple $(\Tt_a)_{a\in A}$.
Further, for every literal $L$, we have that 
$\#_L(\Tt)=\sum_{a\in A} \#_L(\Tt_a)$ and therefore
$\pi(\Tt)=\prod_{a\in A}\pi(\Tt_a)$.
It follows, by the distributive law for tuples, that
\begin{align*}\pi\ext\psi\ &=\ \prod_{a\in A}\pi\ext{\phi(a)}\ =\ 
\prod_{a\in A}\ \ \sum_{\Tt_a\in T(\phi(a))} \pi(\Tt_a)\\
&=\ \sum_{(\Tt_a)_{a\in A}\in T(\psi)}\ \prod_{a\in A} \pi(\Tt_a)\ =\
\sum_{\Tt\in T(\psi)} \pi(\Tt). 
\end{align*}
\end{itemize}
\end{proof}

\section{A Provenance Semiring for First-Order Logic} 
\label{sec:prov-semiring-FOL}

We have claimed in Sect.~\ref{sec:commutative-semirings} that
$\poly{\Nat}{Y}$, the commutative semiring freely generated by a set
$Y$ is used for provenance tracking. The elements of $Y$ label the
information whose propagation we wish to capture in provenance. This
works fine for \emph{positive} database query
languages~\cite{GreenKarTan07} but difference or negation cause problems.
Here we shall use a variation on the idea of polynomials
in order to deal with negated facts in provenance analysis.

We construct a semiring whose elements can be identified with certain
polynomials that describe the provenance of first-order model checking. The
main insight is the use of indeterminates in ``positive-negative pairs''.
We show that the resulting polynomials provide a nicely dual interpretation
for provenance that captures model-checking proofs.
We illustrate this  with a running example. 


\subsection{Dual-Indeterminate Polynomials}
\label{subsec:dual-ind-poly}

Let $X,\nnX$ be two disjoint sets together with a one-to-one
correspondence between $X$ and $nnX$. We denote by $p\in X$ and
$\nnp\in\nnX$ two elements that are in this correspondence.  We refer
to the elements of $X\cup\nnX$ as \textbf{provenance tokens} as they
will be used to label or annotate some of the ``data'', i.e., literals
over some ground values, via the concept of $\Semi$-interpretation that
we defined previously. Indeed, if we fix a finite 
non-empty set $A$ and consider $\LitA(\tau)=\FactsA(\tau)\cup\NegFactsA(\tau)$ then
we shall use $X$ for $\FactsA$ and $\nnX$ for $\NegFactsA$. By
convention, if we annotate $R\bar a$ with the ``positive'' token $p$ 
then the ``negative'' token $\nnp$ can only be
used to annotate $\neg R\bar a$, and vice versa. 
We refer to $p$ and $\nnp$ as \textbf{complementary tokens}.

\begin{definition}
We denote by $\poly{\Nat}{X,\nnX}$ the quotient
of the semiring of polynomials $\poly{\Nat}{X\cup\nnX}$ by the
congruence generated by the equalities $p\cdot\nnp=0$ for
all $p\in X$.\footnote{This is the same as factoring by the ideal generated 
by the polynomials $p\nnp$ for all $p\in X$.} We will call the elements of 
$\poly{\Nat}{X,\nnX}$ \textbf{dual-indeterminate (provenance) polynomials}.
\end{definition}

Observe that two polynomials $\gothp,\gothq\in\poly{\Nat}{X\cup\nnX}$
are congruent if, and only if, they become identical after deleting from each of them
the monomials that contain complementary tokens. Hence, the 
congruence classes in $\poly{\Nat}{X,\nnX}$ are in one-to-one correspondence
with the polynomials in $\poly{\Nat}{X\cup\nnX}$
whose monomials do not contain complementary tokens, so we could have defined
these to be dual-indeterminate polynomials. In particular, we can multiply such
polynomials as usual, provided that we eliminate the monomials with complementary 
tokens afterwards.
The following is the universality property of the semiring of dual-indeterminate
polynomials.
\begin{proposition}
\label{prop:prov-univ}
For any commutative semiring $\Semi$ and for any
$f:X\cup\nnX\rightarrow \Semi$ such that 
$\forall p\in X\,f(p)\cdot f(\nnp)=0$ there exists
a unique semiring homomorphism 
$h:\poly{\Nat}{X,\nnX}\rightarrow \Semi$ 
such that $h(x)=f(x)$ for all $x\in X\cup\nnX$.
\end{proposition}

The dual-indeterminate provenance polynomials serve to track both positive and negative facts about the model throughout a model-checking computation, as we shall illustrate in Sect.~\ref{subsec:ex}.
We note that $\poly{\Nat}{X,\nnX}$ is $+$-positive, but not positive,
since it has divisors of $0$, such as
$$
p\cdot\nnp~=~0,~~~~(p+\nnq)\nnp q ~=~0,~~~~(p\nnq+\nnp q)(pq+\nnp\nnq)~=~0.
$$
However, keeping both $p$ and $\nnp$ around
and even using them in certain ``inconsistent''
$\poly{\Nat}{X, \nnX}$-interpretations 
can be very useful in provenance analysis, as we shall see 
in Sect.~\ref{subsec:rev-ex}, and the subsequent sections.

Finally, we remark that the construction with dual indeterminates can be replicated
for the provenance semirings $\Bool[X]$, $\Sorb(X)$, $\WW(X)$, and $\PosBool(X)$.
For the last one, the result, $\PosBool(X,\nnX)$, corresponds to the usual boolean expressions, in irredundant disjunctive normal form.

\subsection{Provenance Tracking: An Example and a Characterization}
\label{subsec:ex}

We can now consider interpretations into the semiring of dual-indeterminate polynomials.

\begin{definition}
A \textbf{provenance-tracking} interpretation is a 
$\poly{\Nat}{X,\nnX}$-interpretation 
$\pi:\LitA\rightarrow\poly{\Nat}{X,\nnX}$ such that
$\pi(\FactsA)\subseteq X\cup\{0,1\}$
and $\pi(\NegFactsA)\subseteq\nnX\cup\{0,1\}$.
\end{definition}

To track provenance for a given model $\gothA$ we use a provenance-tracking interpretation that is also model-defining, and in fact defines precisely the model $\gothA$. The idea is that if the interpretation annotates a positive or negative fact with
a token, then we wish to track that fact through the model-checking
computation. On the other hand annotating with $0$ or $1$ is done when
we do not track the fact, yet we need to recall whether it holds or not
in the model.

\begin{figure}[h]
\centering
\begin{tikzpicture}
\begin{scope}
  \node[shape=circle,draw=black,very thick] (A) at (0,0) {$a$};
  \node[shape=circle,draw=black,very thick] (B) at (2,1.5) {$b$};
  \node[shape=circle,draw=black,very thick] (C) at (4,0) {$c$};

 \draw[->,very thick] 
      (A) edge[bend right=20] node[right] {$\,p$} (B);
 \draw[->,very thick] 
      (B) edge[bend right=20] node[left] {$q\,$} (C);
 \draw[->,very thick,dashed] (A) edge node[below] {$\nnr$} (C);
 \draw[->,very thick,dashed] 
      (C) edge[bend right=20] node[right] {$\,\nns$} (B);
 \draw[->,very thick] 
      (B) edge[bend right=20] node[left] {$t\,$} (A);
\end{scope}
\end{tikzpicture}
\caption{The model $G$}
\label{fig:modelG}
\end{figure}
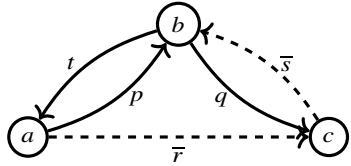

For the following example, the vocabulary of directed graphs consists of one binary predicate $E$ denoting directed edges. 
Consider, over this vocabulary, the following formula and sentence
\[
\domi(x) ~:=~
\forall y\: \bigl( x=y \vee (Exy\wedge\neg Eyx)\bigr),
\qquad
\phi~:=~ \forall x\,\neg\domi(x).
\]
Obviously, $\domi(x)$ says that in a digraph with edge relation $E$ 
the vertex $x$ is ``dominant'' while $\phi$ says
that the digraph does not have a dominant vertex. Consider also, as a model, the digraph $G$ depicted in Figure~\ref{fig:modelG} with vertices $V=\{a,b,c\}$. The finite set $V$ is the universe of ground values for our interpretations.

We adopt the visual convention of representing the edges of the digraph as solid arrows and of labeling them with positive tokens when we wish to track their presence through
model-checking (see $p,q,t$) and with 1 when we are not interested in tracking them (not occurring in this example).
Moreover, we represent with dashed arrows absent edges, but only those
whose absence, however, we also wish to track, by labeling them with negative tokens ($\nnr,\nns$). We extend this to all literals as follows, with the 
provenance-tracking $\poly{\Nat}{X,\nnX}$-interpretation
$\beta:\LitV\rightarrow X\cup\nnX\cup\{0,1\}$ defined by
\[ \beta(L) ~=~ \begin{cases}
           p  & \mathrm{if~} L=Eab\\
           0  & \mathrm{if~} L=\neg Eab\\
           q  & \mathrm{if~} L=Ebc\\
           0  & \mathrm{if~} L=\neg Ebc\\
           0  & \mathrm{if~} L=Eac \\
        \nnr  & \mathrm{if~} L=\neg Eac
            \end{cases}
 \quad\text{ and }\quad\beta(L)~=~\begin{cases}
           0  & \mathrm{if~} L= Ecb\\
        \nns  & \mathrm{if~} L=\neg Ecb\\
           t  & \mathrm{if~} L= Eba\\
           0  & \mathrm{if~} L=\neg Eba\\
           0  & \mathrm{for~the~other~positive~facts}\\
           1  & \mathrm{for~the~other~negative~facts.}
            \end{cases}
\]

For example, $\beta(Eca)=\beta(Ecc)=\ldots=0$ and 
$\beta(\neg Eca)=\beta(\neg Ecc)=\ldots=1$. 
Note that $\beta$ is \ModDef\ in the sense of
of Sect.~\ref{def:int-model-defining}
and that the model it defines is precisely $G$.

The assumptions made in the definition of $\beta$ 
indicate that we choose to track positive facts
like $Ebc$ and negative facts like $\neg Eab$, etc., as they are
used in establishing the truth of some sentence in $G$.
They also indicate that we accept, and thus do not
track, the absence of the other potential edges such as $Eca$.
We think of data annotated with $0$ as being ``forget-about-it'' absent and of 
data annotated with 1 as ``available for free'' present.

Clearly, $G\models\phi:=\forall x\,\neg\domi(x)$. But how can we justify this in terms
of the facts, negative or positive, that hold in the model?
By computing the semantics of $\phi$
under the interpretation $\beta$ we will obtain \emph{provenance information}
for the result that $G\models\phi$. Clearly
$$
\nnf(\phi)~=~ 
\forall x\,\exists y\: \bigl( x\neq y \wedge (\neg Exy\vee Eyx)\bigr)
$$
and therefore
$$
\beta\sem{\phi}{} ~=~ \beta\sem{\nnf(\phi)}{} ~=~
(\nnr+t)\cdot p\cdot(1+q+\nns) ~=~ p\nnr+pt+pq\nnr+pqt+p\nnr\nns+p\nns t.
$$

Each of the monomials of the polynomial $\beta\sem{\phi}{}$
has coefficient 1 and all the exponents are 1. This is certainly not the case
in general. 
For instance, we could have changed the provenance assumptions
$\beta$ to have 1 instead of $r$ (and therefore 0 instead of $\nnr$).
In addition suppose that the potential presence/absence of
all three edges $Eab,Ecb$ and $Eba$ has the \emph{same} provenance
(e.g., same data source) so we choose to track their presence/absence
with the same tokens $p/\nnp$. \val{Not model-defining!}
This results in $\beta\sem{\phi}{}=(\nnp+p)(2p+\nnq+\nnp)(2+q+\nnp)
=\ldots+4p^2+\ldots+2\nnp\nnq+\ldots+\nnp^3+\ldots$.
In fact, it is possible to show that any polynomial
can be computed as some provenance, with suitable choices of sentence,
model, and interpretation.

Each of the monomials obtained with a provenance-tracking interpretation corresponds to a 
different (model-checking) proof tree of $\phi$ from the literals
described by the monomial. 
We illustrate this with the proof tree 
corresponding to another monomial, $p\nnr\nns$,
using the following formula abbreviations:
\begin{eqnarray*}
\denydom(x,y) &:=& \bigl(x\neq y \wedge (\neg Exy\vee Eyx)\bigr)
~~~~~~~~~~~~y~\mathrm{denies~dominance~of}~x\\
\\
\notdom(x)   &:=& 
       \exists y\: \bigl( x\neq y \wedge (\neg Exy\vee Eyx)\bigr)
~~~~~~~~~~~x~\mathrm{is~not~dominant}\\
\\
\noVdom      &:=& 
     \forall x\,\exists y\: \bigl(x\neq y \wedge (\neg Exy\vee Eyx)\bigr)
~~~~~~~\mathrm{no~vertex~is~dominant}\\
\end{eqnarray*}

\vspace*{-5mm}

Further, we abbreviate 
$\neg Exy\vee Eyx$ by $\theta(x,y)$. With these, the proof tree corresponding to $p\nnr\nns$ is:

\bigskip

\EnableBpAbbreviations

\AXC{$a\neq c$}
\AXC{$\neg Eac\ [\nnr]$}
\UIC{$\theta(a,c)$}
\BIC{$\denydom(a,c)$}
\UIC{$\notdom(a)$}
\AXC{$b\neq a$}
\AXC{$Eab\ [p]$}
\UIC{$\theta(b,a)$}
\BIC{$\denydom(b,a)$}
\UIC{$\notdom(b)$}
\AXC{$c\neq b$}
\AXC{$\neg Ecb\ [\nns]$}
\UIC{$\theta(c,b)$}
\BIC{$\denydom(c,b)$}
\UIC{$\notdom(c)$}
\TIC{$\noVdom$}
\DP\\

\cutout{With these, the proof tree corresponding to $p\nnr\nns$ is:

\bigskip

\EnableBpAbbreviations

\noindent
\AXC{$a\neq c$}
\AXC{$\neg Eac~~~[\nnr]$}
\UIC{$\neg Eac\vee Eca$}
\BIC{$\denydom(a,c)$}
\UIC{$\notdom(a)$}
\AXC{$b\neq a$}
\AXC{$Eab~~~[p]$}
\UIC{$\neg Eba\vee Eab$}
\BIC{$\denydom(b,a)$}
\UIC{$\notdom(b)$}
\AXC{$c\neq b$}
\AXC{$\neg Ecb~~~[\nns]$}
\UIC{$\neg Ecb\vee Ebc$}
\BIC{$\denydom(c,b)$}
\UIC{$\notdom(c)$}
\TIC{$\noVdom$}
\DP\\
}

The formulae on the leaves of the tree are accompanied by the tokens that 
$\beta$ annotates them with.

The following proposition, which is an immediate consequence of Theorem~\ref{thm:prooftrees}, summarizes the situation.

\begin{proposition}
\label{prop:all-proofs-mod-def}
Let $\beta:\LitA\rightarrow\poly{\Nat}{X,\nnX}$ be a provenance-tracking
\ModDef\ interpretation,
and let $\phi$ be an \FO-sentence. 
Then, the dual-indeterminate polynomial
$\beta\sem{\phi}{}$ describes \emph{all} the proof
trees that verify $\phi$ using premises from among the literals
that $\beta$ maps to provenance tokens or to 1 (i.e., from the
literals that hold in $\Abeta$).
Specifically, each monomial $m\,x_1^{m_1}\cdots x_k^{m_k}$ 
corresponds to $m$ distinct proof trees
that use $m_1$ times a literal that $\beta$ annotates by $x_1$, \ldots, 
and $m_k$ times a literal annotated by $x_k$, as well
as any number of the literals annotated with 1.
In particular, $\beta\sem{\phi}{}\neq0$ if, and only if, some proof tree exists, and if, and only if,
$\Abeta\models\phi$.
\end{proposition}

Note that since $\poly{\Nat}{X,\nnX}$ is not positive this proposition
does not follow from Proposition~\ref{prop:strong}.
(Nor does this contradict Proposition~\ref{prop:refi} because 
provenance-tracking interpretations have a special form.)
Nonetheless, albeit not positive,
$\poly{\Nat}{X,\nnX}$ has many remarkable properties and
this proposition is a corollary of a more general one that we shall 
state in Sect.~\ref{subsec:prop-prov}.

\subsection{The size of provenance}
\label{subsec:size}

In this section we explore the size of provenance as a complexity measure. In~Sect.~\ref{subsec:dual-ind-poly} we have introduced the dual-indeterminate polynomials as formal embodiments of the provenance of model checking assertions $\gothA\models\phi$. However, syntactically, these polynomials are normal forms of expressions built from tokens with the operations $+,\cdot,0,1$. Many such expressions are equationally equivalent under the laws of semirings. Thus, we can think of \emph{representing} provenance as polynomials (normal forms), or, more economically, parenthesized expressions, or even as circuits in which common subexpressions are shared. A simple induction on the structure of formulas yields:

\begin{proposition}
Provenance of $\gothA\models\phi$ computed as parenthesized expression has size polynomial in $|\gothA|$ and exponential in $|\phi|$.
\end{proposition}

On the other hand, converting provenance to polynomial normal form can involve a combinatorial explosion in the number of monomials (although the size of each monomial remains polynomial because there are only polynomially many facts). Indeed, consider the universe $\{a_1,\ldots,a_n\}$, the sentence (a trivial tautology) $\forall x (Rx\lor\neg Rx)$ and the interpretation $\pi(Ra_i)=p_i, \pi(\neg Ra_i)=\nnp_i$. As a parenthesized expression, the model checking provenance is $(p_1+\nnp_1)\cdots(p_n+\nnp_n)$
which has size $O(n)$ while in polynomial normal form it has size $\Omega(2^n)$.

This is due to the presence of the universal quantifiers. Indeed, an examination of the proof of the previous proposition yields:

\begin{proposition}
If $\phi$ is an existential sentence in NNF then the dual-indeterminate polynomial corresponding to checking it in a model $\gothA$ has only polynomially (in $|\gothA|$) many monomials.
\end{proposition}

\subsection{Explanations Obtained from Provenance}
\label{subsec:expl}

In~Sect.~\ref{subsec:ex} we have calculated the provenance of the sentence $\phi$ that asserts ``there is no dominant vertex'',
\[
\phi~:=~ \forall x\,\neg\domi(x) \quad\text{ where }\quad
\domi(x) ~:=~
\forall y\: \bigl( x=y \vee (Exy\wedge\neg Eyx)\bigr),
\]
being checked in the model and interpretation in ~Fig.~\ref{fig:modelG}, resulting in the polynomial
$p\nnr+pt+pq\nnr+pqt+p\nnr\nns+p\nns t$. As we have explained, each monomial of this polynomial corresponds to a proof tree for model-checking.

As a practical matter, we can think of each of the monomials of this polynomial as an \emph{alternative explanation}.
For example, the monomial $p\nnr\nns$ (shown to correspond to a proof tree illustrated above) gives an explanation that can be stated informally as follows: there is no dominant vertex because of the presence of an edge from $a$ to $b$ together with the absence of edges from
$a$ to $c$ and from $c$ to $b$. 

Note that if we examine the proof tree we can obtain more information: the presence of $Eab$ is used to show that $b$ is not dominant, the absence of $Eac$ shows that $a$ is not dominant, and the absence of $Ecb$ shows that $c$ is not dominant; therefore none of the vertices is dominant. 

Another example that helps underlining the difference between monomials and proof trees is given by $p\nnr$.
As an explanation, this monomial is a strict part of the explanation supplied by $p\nnr\nns$. However,
$p\nnr$ corresponds to a proof tree of $\phi$ 
in which the presence of $Eab$ and the absence of $Eac$ are used as in the proof tree for $p\nnr\nns$ above, yet, this tree uses the absence of  $Eca$, which is \emph{accepted without 
tracking}---it has provenance 1---to show that $c$ is not dominant. Overall for $\phi$ we could stick for practical purposes to just minimal explanations, which here are just $p\nnr$ and $pt$.

\subsection{From Provenance to Confidence}
\label{subsec:conf}

Recall from Sect.~\ref{sec:commutative-semirings} the Viterbi semiring
$\Vit$. We think of the elements of $\Vit$ as confidence scores. 
Going back to the example in Sect.\ref{subsec:ex}, and assuming
specific confidence scores for the literals that $G$ makes true,
and that we track, we wish to compute a confidence score for
$G\models\phi$.

Specifically, consider the $\Vit$-interpretation 
$\gamma:\LitV\rightarrow[0,1]$ defined by
\begin{itemize}
\item $\gamma(Eab)=\gamma(Ebc)=0.9,
\gamma(Eba)=0.2$,  and $\gamma(Euv)=0$  for any \emph{other} positive fact $Euv$,
\item $\gamma(\neg Euv)=0$ whenever $\gamma(Euv)\neq 0$,
\item $\gamma(\neg Eac)=\gamma(\neg Ecb)=0.6$, and
\item $\gamma(\neg Euv)=1$ for any \emph{other} negative fact with $\gamma(Euv)=0$.
\end{itemize}

With this we could use Definition~\ref{def:sem}
to compute $\gamma\sem{\phi}{}\in[0,1]$, which is the desired
confidence score.

However, since we have already computed in Sect.~\ref{subsec:ex} the provenance
$\beta\sem{\phi}{}$ we can take advantage of the Fundamental Property
(Proposition~\ref{prop:hom}) via a homomorphism whose existence is
guaranteed by Proposition~\ref{prop:prov-univ}.

We define $f:X\cup\nnX\rightarrow[0,1]$ by
\begin{itemize}
\item $f(p)=f(q)=0.9$, $f(t)=0.2$, and $f(x)=0$ for $x\not\in\{p,q,t\}$,
\item $f(\nnx)=0$ for $x\in\{p,q,t\}$,
\item $f(\nnr)=f(\nns)=0.6$, and
\item $f(\nnx)=1$ for  $\nnx\not\in\{\nnp,\nnq,\nnt,\nnr,\nns\}$.
\end{itemize}

The condition on $f$ in 
Proposition~\ref{prop:prov-univ} is satisfied, hence $f$ can be extended
to a homomorphism $h:\poly{\Nat}{X,\nnX}\rightarrow\Vit$. From the definition
of $f$ we have $h\circ\beta=\gamma$. By the Fundamental Property
$$
\gamma\sem{\phi}{}~=~ h(\beta\sem{\phi}{}).
$$
Hence the score we wish to compute can be obtained by applying the homomorphism
$h$ to the dual polynomial $\beta\sem{\phi}{}=p\nnr+pt+pq\nnr+pqt+p\nnr\nns+p\nns t$. It is easier to use the 
factored form of $\beta\sem{\phi}{}$:
$$
h(p(\nnr+t)(1+q+\nns)) ~=~ 0.9\cdot\max(0.6,\;0.2)\cdot\max(1,\;0.9,\;0.6)
~=~ 0.54.
$$
In general, confidence calculation may be only one of the analyses that we 
wish to perform. When these analyses are based on semiring calculations
we can compute the provenance just once and then evaluate it in multiple
semirings and under multiple valuations, by virtue of the Fundamental Property.

\subsection{Detailed Provenance Analysis: Top-Secret Proofs}
\label{subsec:clear}

We describe here another kind of provenance analysis that we can perform 
in conjunction with interpretation in various semirings.
Recall from Sect.~\ref{sec:commutative-semirings} the access control
semiring $\Access$. Its elements are interpreted as \emph{clearance levels}
$0 < \Tsec < \Sec < \Cnf < \Pub=1$. For example, administrators would assign 
clearance levels
to the different items in the input data. The resulting clearance
level for the output of a computation determines which users get to
access that output.  In the context of this paper there would be an
assignment of clearance levels to literals. 

Going back to the example in Sect.~\ref{subsec:ex}, 
consider the $\Access$-interpretation 
$\alpha:\Lit_V(\tau)\rightarrow\Access$ defined by setting
$$
\alpha(Eab)=\alpha(Ebc)=\alpha(Eba)=\Pub,
~~~~~~~~~~~~~
\alpha(\neg Eac)=\alpha(\neg Ecb)=\Tsec,
$$
and $\alpha(Euv)=0$ for any \emph{other} positive fact,
$\alpha(\neg Euv)=\Pub$ for the corresponding negative fact.

As in Sect.~\ref{subsec:conf} we have
$\alpha\sem{\phi}{}=h(p\nnr+pt+pq\nnr+pqt+p\nnr\nns+p\nns t)$,
where $h$ is the unique homomorphism $\poly{\Nat}{X,\nnX}\rightarrow\Access$
such that $h(p)=h(q)=h(t)=\Pub$,
$h(\nnr)=h(\nns)=\Tsec$, and otherwise equals $0$ on the rest of $X$
and equals $\Pub$ on the rest of $\nnX$. 

We can see that $\alpha\sem{\phi}{}=\Pub$ but we can also perform a
more detailed analysis in which we can associate clearance levels to
individual proof trees.
Thus, while it will be
publicly known that $G\models\phi$, those with top-secret
clearance can know that also $p\nnr$ describes a proof of the
assertion $G\models\phi$. This may become relevant if we have
particularly high confidence (as described above in Sect.~\ref{subsec:conf})
in the literals that $p$ and $\nnr$ annotate, that is, in the presence
of the edge from $a$ to $b$ and in the absence of an edge from $a$ to $c$.

\section{Reverse Provenance Analysis}
\label{sec:reverse}

There are limitations to what we can do with the provenance of a
model-checking assertion $\gothA\models\phi$ under an interpretation that defines the model $\gothA$.
It is even more interesting to consider provenance-tracking interpretations
that allow us to \emph{choose}, from among multiple models, the ones
that fulfill various desiderata.

\begin{figure}[h]
\centering
\begin{minipage}[t]{.5\textwidth}
\centering
\begin{tikzpicture}
\begin{scope}
  \node[shape=circle,draw=black,very thick] (A) at (0,0) {$a$};
  \node[shape=circle,draw=black,very thick] (B) at (2,1.5) {$b$};
  \node[shape=circle,draw=black,very thick] (C) at (4,0) {$c$};

 \draw[->,very thick,dotted] 
      (A) edge[bend right=20] node[right] {$p,\nnp$} (B);
 \draw[->,very thick,dotted] 
      (B) edge[bend right=20] node[left] {$q,\nnq\,$} (C);
 \draw[->,very thick,dotted] (A) edge node[below] {$r,\nnr$} (C);
 \draw[->,very thick,dotted] 
      (C) edge[bend right=20] node[right] {$s,\nns$} (B);
 \draw[->,very thick,dotted] 
      (B) edge[bend right=20] node[left] {$t,\nnt$} (A);
\end{scope}
\end{tikzpicture}
\vspace*{-3mm}
\caption{Provenance tracking assumptions}
\label{fig:assump}
\end{minipage}%
~~~~
\begin{minipage}[t]{.5\textwidth}
\centering
\begin{tikzpicture}
\begin{scope}
  \node[shape=circle,draw=black,very thick] (A) at (0,0) {$a$};
  \node[shape=circle,draw=black,very thick] (B) at (2,1.5) {$b$};
  \node[shape=circle,draw=black,very thick] (C) at (4,0) {$c$};

 \draw[->,very thick] 
      (A) edge[bend right=20] node[right] {$\,p$} (B);
 \draw[->,very thick] 
      (B) edge[bend right=20] node[left] {$q\,$} (C);
 \draw[->,very thick] (A) edge node[below] {$r$} (C);
 \draw[->,very thick] 
      (C) edge[bend right=20] node[right] {$\,s$} (B);
 \draw[->,very thick] 
      (B) edge[bend right=20] node[left] {$t\,$} (A);
\end{scope}
\end{tikzpicture}
\vspace*{-3mm}
\caption{The model $\calF$}
\label{fig:modelF}
\end{minipage}%
\end{figure}

\subsection{A Reverse Analysis Example}
\label{subsec:rev-ex}

Let $V=\{a,b,c\}$ be a set of ground values. As before, these will eventually
play the role of the vertices of a digraph.
However, we do not yet specify
a set of edges, i.e., we do not specify a finite model with
universe $V$. Instead, as illustrated by the dotted edges in 
Figure~\ref{fig:assump}, we supply a set of provenance tokens 
$X=\{p,q,r,s,t\}$ that correspond
to the \emph{potential presence} of some edges that we wish to track.
Therefore, $\nnX=\{\nnp,\nnq,\nnr,\nns,\nnt\}$ are the provenance tokens 
allowing us to track the \emph{potential absence} of the same edges.
These \textbf{provenance tracking assumptions} can be formalized via a 
provenance-tracking $\poly{\Nat}{X,\nnX}$-interpretation.

Define $\pi:\Lit_V(\{E\})\rightarrow X\cup\nnX\cup\{0,1\}$ by
\[
\pi(L) ~=~ \begin{cases}
           p  & \mathrm{if~} L=Eab\\
        \nnp  & \mathrm{if~} L=\neg Eab\\
           q  & \mathrm{if~} L=Ebc\\
        \nnq  & \mathrm{if~} L=\neg Ebc\\
           r  & \mathrm{if~} L=Eac \\
        \nnr  & \mathrm{if~} L=\neg Eac
            \end{cases}
\quad\text{ and }\quad\pi(L)=  \begin{cases}
           s  & \mathrm{if~} L= Ecb\\
        \nns  & \mathrm{if~} L=\neg Ecb\\
           t  & \mathrm{if~} L= Eba\\
        \nnt  & \mathrm{if~} L=\neg Eba\\
           0  & \mathrm{for~the~other~positive~facts}\\
           1  & \mathrm{for~the~other~negative~facts.}
            \end{cases}
\]

For example, $\pi(Eca)=\pi(Ecc)=\ldots=0$ and
$\pi(\neg Eca)=\pi(\neg Ecc)=\ldots=1$. 
This particular interpretation does not feature a positive fact annotated
with 1 but we could have just as well had $\pi(Eab)=1$ and 
$\pi(\neg Eab)=0$ if we chose to assume that edge without tracking it.

Note that $\pi$ is not \ModDef\ (in the sense of
Definition~\ref{def:int-model-defining}), i.e., it does not correspond
to any \emph{single} model. 
As we shall see, this is not a bug but a feature, as it
will allow us to consider, under the given provenance assumptions,
multiple models that can satisfy a sentence.

Now we compute the semantics of the sentence 
$\phi\equiv~ \forall x\,\neg\domi(x)$
from Sect.~\ref{subsec:ex},
under this interpretation and we obtain 
$$
\pi\sem{\phi}{} ~=~ 
(\nnp+\nnr+t)\cdot(p+\nnq+s+\nnt)\cdot(1+q+r+\nns).
$$

If we multiply the three sums and  apply
$p\nnp=q\nnq=r\nnr=s\nns=0$ we get a polynomial with 
$48-4-3-3-4=34$ monomials (the reader shall be spared the
pleasure of admiring it). 
\erich{changed trouble to pleasure :-)}
As in Sect.~\ref{subsec:ex},
each of these monomials has coefficient 1
and (as shown in Sect.~\ref{subsec:prop-prov})
each corresponds to a different proof tree of $\phi$ from the literals
described by the monomial.

For example, the monomial $pqt$ corresponds to a proof tree of $\phi$ 
in which the fact
$Eba$ is used to deny the dominance of $a$, the fact
$Eab$ is used to deny the dominance of $b$, and the fact
$Ebc$ is used to deny the dominance of $c$. Recalling the
notations from Sect.~\ref{subsec:ex}, note that the same monomial
is part of the dual polynomial $\beta\sem{\phi}{}$ and that
the same proof tree justifies $G\models\phi$. Note also
that setting $r=s=\nnp=\nnq=\nnt=0$ in the definition
of $\pi$ gives the definition of $\beta$. Doing the same in 
$\pi\sem{\phi}{}$ gives
$$
(0+\nnr+t)\cdot(p+0+0+0)\cdot(1+q+0+\nns) 
~=~
(\nnr+t)\cdot p\cdot(1+q+\nns),
$$ 
which is the same as the polynomial $\beta\sem{\phi}{}$ obtained with the
\ModDef\ interpretation $\beta$ which corresponds to the model $G$. In
this sense, $\pi$ is a ``generalization'' of $\beta$, or, 
$\beta$ can be obtained by \emph{specializing} $\pi$.
All this will be made precise in full generality
in Sect.~\ref{subsec:prop-prov} while here we explore two other interesting
specializations of $\pi$.

One of the monomials in $\pi\sem{\phi}{}$ is $\nnp\nnq$.  This means
that we can find a specialization of $\pi$ that is \ModDef\ and that defines,
in fact, 
a model with \emph{no} positive information, namely the digraph with
vertices $V$ and no edges.
Hence, denoting with $\gothE$
this no-edge model, we have $\gothE\models\phi$.
How many proof trees verify that $\gothE\models\phi$?
The specialization of $\pi$ that we are after (let's call it $\beta_1$)
corresponds to
setting $p=q=r=s=t=0$ in $\pi$ and in $\pi\sem{\phi}{}$. This gives
$$
\beta_1\sem{\phi}{} ~=~ 
(\nnp+\nnr)\cdot(\nnq+\nnt)\cdot(1+\nns),
$$
which is a polynomial with 8 monomials, each with coefficient 1. It follows
that there are 8 distinct proof trees for $\gothE\models\phi$.

One can also figure out that $pqt, prt, qst, rst$ are among the
monomials in $\pi\sem{\phi}{}$. This means that we can find another
specialization of $\pi$ (let's call it $\beta_2$) that is also \ModDef\ and that defines a model with
\emph{maximum} positive information (allowed by $\pi$), namely the 
digraph with
vertices $V$ and edges $Eab,Ebc,Eac,Ecb$ and $Eba$.
Let's denote with $\calF$
this all-allowed-edges model (see Figure~\ref{fig:modelF}). 
How many proof trees verify that $\calF\models\phi$?
The specialization $\beta_2$ that we look for here 
corresponds to setting $\nnp=\nnq=\nnr=\nns=\nnt=0$. This gives
$$
\beta_2\sem{\phi}{} ~=~ 
t\cdot(p+s)\cdot(1+q+r),
$$
which is a polynomial with 6 monomials, each with coefficient 1, 
hence there are 6 proof trees
for this.

Finally, we also wish to consider for this example 
the provenance of the \emph{negation}
of the sentence $\phi$ considered above, i.e., the sentence
$\neg\phi$ that says that the digraph \emph{has} a dominant vertex:
$$
\neg \phi~=~ \neg \forall x\,\neg\domi(x).
$$
Since $\domi(x) = 
\forall y\: \bigl(x=y \vee (Exy\wedge\neg Eyx)\bigr)
$
is already in NNF\val{Are we consistent in using the NNF acronym?}, we have $\nnf(\neg\phi)=\exists x\,\domi(x)$.
We compute the semantics of this sentence under the same interpretation:
$$
\pi\sem{\neg\phi}{}~=~ pr\nnt + \nnp q\nns t + 0\cdot\nnq\nnr s
                   ~=~ pr\nnt + \nnp q\nns t.
$$
Thus, under the provenance tracking assumptions we have made,
there are only two proof trees for $\neg\phi$.

The attentive reader may have also noticed a certain ``duality'' between $\pi\sem{\neg\phi}{}$ above
(specifically, the expression that includes $0\cdot\nnq\nnr s$)
and $\pi\sem{\phi}{}$ given earlier, if we think of + as dual to $\cdot$ and
$p$ as dual to $\nnp$, etc. Indeed, this duality can be made precise 
(before dealing with complementary tokens, that is, in $\poly{\Nat}{X\cup\nnX}$ rather than
in $\poly{\Nat}{X,\nnX}$) and we do that in~Sect.~\ref{subsec:prop-prov}.

\subsection{Properties of Provenance} 
\label{subsec:prop-prov}

The interpretation exhibited
in Sect.~\ref{subsec:rev-ex} belongs to a class that merits its own
definition.

\begin{definition} A provenance-tracking interpretation
$\pi:\LitA(\tau)\rightarrow\poly{\Nat}{X,\nnX}$
is said to be \bfModComp\
if for each fact $R\bar a$ one of the
following (mutually exclusive) three properties holds:
\begin{enumerate}
\item $\exists z\in X$
s.t $\pi(R\bar a)=z$ and $\pi(\neg R\bar a)=\nnz$,
or
\item $\pi(R\bar a)=0$ and $\pi(\neg R\bar a)=1$, or
\item $\pi(R\bar a)=1$ and $\pi(\neg R\bar a)=0$
\end{enumerate}
\end{definition}

As promised, we state a more powerful
version of Proposition~\ref{prop:all-proofs-mod-def}
(which was about provenance-tracking \ModDef\ interpretations).

\begin{theorem}
\label{thm:modelcompatible}
Let $\pi:\LitA(\tau)\rightarrow\poly{\Nat}{X,\nnX}$ be a \ModComp\
interpretation and let $\psi$ be sentence in $\FO(\tau)$. Then,
$\pi\sem{\psi}{}$ describes \emph{all} the proof
trees that verify $\phi$ using premises from among the literals that
$\pi$ maps to provenance tokens or to 1.  Specifically, each monomial
$m\,x_1^{m_1}\cdots x_k^{m_k}$ corresponds to $m$ distinct proof trees
that use $m_1$ times a literal annotated by $x_1$, \ldots, and $m_k$ times
a literal annotated by $x_k$, where $x_1,\ldots,x_k\in X\cup\nnX$.
In particular, when $\pi\sem{\psi}{}=0$ no proof tree exists.
\end{theorem}

Again, the proof is a direct consequence of Theorem~\ref{thm:prooftrees}.
Note that, in this context, an evaluation tree 
$\Tt$ for a \ModComp\ interpretation $\pi:\LitA(\tau)\rightarrow\poly{\Nat}{X,\nnX}$ has the valuation
\[ \pi(\Tt):=\prod_{x\in X\cup\nnX} x^{\#_x(\Tt)},\]
where $\#_x(\Tt)$ is the number of leaves of $\Tt$ that
$\pi$ maps to $x$. Further, $\pi(\Tt)=0$ if $\Tt$ contains 
leaves mapped to complementary provenance tokens.

\begin{corollary}
\label{cor:count}
Let $\pi$ be a \ModComp\ interpretation. Then, the sum of
the monomial coefficients in $\pi\sem{\phi}{}$ counts 
the number of proof trees that verify $\phi$ using premises 
from among the literals that $\pi$ maps to provenance tokens or to 1.
The same count can be obtained from an $\Nat$-interpretation 
as $(h\circ\pi)\sem{\phi}{}\in\Nat$ where 
$h:X\cup\nnX\cup\{0,1\}\rightarrow\Nat$ is defined by $h(0)=0$ and
$h(p)=h(\nnp)= h(1)=1$.
\end{corollary}

A \ModComp\ interpretation may allow the tracking of both a literal
and its negation. Therefore, \ModComp\ interpretations are not 
\ModDef\ unless they do not make use of provenance tokens at all 
(in which case they are essentially canonical truth interpretations). 
Hence, 
Proposition~\ref{prop:all-proofs-mod-def} is not a simple particular case of 
Theorem~\ref{thm:modelcompatible}. Nonetheless, we shall see how
\ModDef\ interpretations can be seen as \emph{specializations} of \ModComp\ 
interpretations with respect to models that ``agree'' 
(i.e., are \emph{compatible}) with them, as defined below. 

\begin{definition}
Let $\pi:\LitA(\tau)\rightarrow\poly{\Nat}{X,\nnX}$ be a \ModComp\
interpretation and let $\gothA$
be a model with universe $A$. We say that $\gothA$
is \textbf{compatible} with $\pi$
if $\gothA\models L$ for any literal $L$ such that $\pi(L)=1$. 
Further, let $\Modpi:=\{\gothA\mid\gothA\mathrm{~is~compatible~with~}\pi\}$.
\end{definition}

For instance, the models shown in 
Figures~\ref{fig:modelG} and~\ref{fig:modelF}
are compatible with the interpretation
defined in Sect.~\ref{subsec:rev-ex}.

Now we can talk about satisfiability and validity \emph{restricted to the
class of models that agree with the provenance tracking assumptions}
made by an interpretation.

\begin{corollary}[to Theorem~\ref{thm:modelcompatible}]
\label{cor:sat}
Let $\pi:\LitA(\tau)\rightarrow\poly{\Nat}{X,\nnX}$ be a \ModComp\
interpretation and let $\phi$ be a first-order sentence. 
Then, 
\begin{itemize}
    \item $\phi$ is $\Modpi$-satisfiable if, and only if,  $\pi\sem{\phi}{}\neq0$,
    \item $\phi$ is $\Modpi$-valid if, and only if,  $\pi\sem{\neg\phi}{}=0$.
\end{itemize}
\end{corollary}

This is not finite satisfiability (shown undecidable by Trakhtenbrot),
of course.
Even if we map every possible literal to a different provenance token
we only decide satisfiability in a model with exactly $|A|$ elements,
which is easily in NP (without talking about provenance).

\begin{example}
\label{ex:tautology}
With the same (digraph) vocabulary as in Sect.~\ref{subsec:ex} 
and~Sect.~\ref{subsec:rev-ex} consider the sentence
$$
\psi := \exists x\,\forall y\,Exy\,\rightarrow\,
            \forall y\,\exists x\,Exy.
$$
This is a well-known tautology (holding in all models, not just in finite
ones). Obviously,
$\nnf(\neg\psi) = \exists x\,\forall y\,Exy\,\wedge\,\exists y\,\forall x\,\neg Exy$. 
Now consider $V=\{a,b\}$ and a truth-compatible interpretation
$\pi$ that annotates $Eab, Eba, Eaa, Ebb$ with $p,q,r,s$ 
respectively,
and the corresponding negated facts with $\nnp,\nnq,\nnr,\nns$. Then
$$
\pi\sem{\neg\psi}~=~(pr+qs)(\nnq\nnr+\nnp\nns)=0,
$$
verifying that $\psi$ is $\Modpi$-valid. 
\end{example}

From the provenance analysis of (provenance-restricted) 
validity/satisfiability that is enabled by Corollary~\ref{cor:sat}
we can obtain a provenance analysis of model checking, for each
model of a given sentence, as follows.

\begin{definition}
\label{def:spec}
Let $\pi$ be \ModComp\ and let $\gothA\in\Modpi$.
The \textbf{specialization} of $\pi$ with respect to $\gothA$
is the $\poly{\Nat}{X,\nnX}$-interpretation
$\refi{\pi}{\gothA}:\LitA\rightarrow\poly{\Nat}{X,\nnX}$ defined by
$$
\refi{\pi}{\gothA}(L) ~=~ \begin{cases}
                          \pi(L) & \mathrm{if~}\gothA\models L\\
                          0      & \mathrm{otherwise.}
                          \end{cases}
$$
\end{definition}

Note that $\refi{\pi}{\gothA}$ is always \ModDef\ and the model it defines
is, of course, $\gothA$. Note also, that for any sentence $\phi$ the dual-indeterminate  polynomial 
$\refi{\pi}{\gothA}\sem{\phi}{}$ is obtained by replacing in 
$\pi\sem{\phi}{}$ the tokens $\pi(L)$ with $0$ for all literals $L$ such that $\gothA\not\models L$.

The \ModDef\ interpretation $\beta$ in Sect.~\ref{subsec:ex} is the specialization
with respect to the model $G$ of the \ModComp\ interpretation $\pi$
in Sect.~\ref{subsec:rev-ex}, $\beta=\refi{\pi}{G}$. 
Other specializations of $\pi$ are given
in Sect.~\ref{subsec:rev-ex}. The next corollary implies
Proposition~\ref{prop:all-proofs-mod-def}.

\begin{corollary}[to Theorem~\ref{thm:modelcompatible}]
\label{cor:model-check}
Let $\pi:\LitA(\tau)\rightarrow\poly{\Nat}{X,\nnX}$ be a \ModComp\
interpretation, let $\gothA$ be a $\tau$-structure that is compatible with $\pi$, and let 
$\phi$ be a first-order sentence such that $\gothA\models\phi$ (hence, by
Corollary~\ref{cor:sat}, $\pi\sem{\phi}{}\neq0$).

Then, $\refi{\pi}{\gothA}\sem{\phi}{}\neq0$ and
every monomial in $\refi{\pi}{\gothA}\sem{\phi}{}$
also appears in $\pi\sem{\phi}{}$, with the same coefficient.

Moreover, $\refi{\pi}{\gothA}\sem{\phi}{}$ describes 
\emph{all} the proof trees that verify $\gothA\models\phi$. In particular,
the sum of all the monomial coefficients in $\refi{\pi}{\gothA}\sem{\phi}{}$
counts the number of distinct such proof trees 
(as in Corollary~\ref{cor:count}, the same count can be obtained from an
$\Nat$-interpretation).
\end{corollary}

While $\refi{\pi}{\gothA}\sem{\phi}{}$ analyzes the provenance
of checking in a specific model, the more general
$\pi\sem{\phi}{}$ allows for a form \emph{reverse analysis}.
Indeed, to each monomial $m\,x_1^{m_1}\cdots x_k^{m_k}$ in
$\pi\sem{\phi}{}\neq0$ we can associate a model from $\Modpi$ that
makes true the literals that are annotated by $x_1,\ldots,x_k$
(and possibly more literals)
and, as we have seen, every model $\gothA\in\Modpi$
such that $\gothA\models\phi$ can be obtained this way.

\begin{example}[Example~\ref{ex:tautology} cont'd]
\label{ex:tautology-two}
Let us also compute the provenance of the tautology $\psi$ itself:
$$
\pi\sem{\psi}{}=(\nnp+\nnr)(\nnq+\nns)+(q+r)(p+s).
$$
Here $\Modpi$ consists of all possible structures with universe $\{a,b\}$
and, for any such $\gothA$, the model-refinement $\refi{\pi}{\gothA}$
sets to 0 exactly one of the two tokens in a complementary pair.
No matter how this is done, observe that 
$\refi{\pi}{\gothA}\sem{\tau}{}\neq0$.
\end{example}

\textbf{Dualizing dual-indeterminate polynomials. }This is best defined on \emph{expressions} over the semiring 
$\poly{\Nat}{X\cup\nnX}$ and then separately discussing the effect of factoring by $p\cdot\nnp=0$. Dualizing
is defined inductively as follows:
$$
\begin{array}{rcl@{\qquad}rcl}
\mathrm{dual}(e_1+e_2) & := & e_1\cdot e_2 &
\mathrm{dual}(e_1\cdot e_2) & := & e_1+e_2\\
\mathrm{dual}(p) & := & \nnp &
\mathrm{dual}(\nnp) & := & p\\
\mathrm{dual}(0) & := & 1  &
\mathrm{dual}(1) & := & 0\\
\end{array}
$$
Note that dualization is self-inverse.
\begin{proposition}Let 
$\pi:\LitA\rightarrow\poly{\Nat}{X\cup\nnX}$ 
be a model-compatible interpretation. Then, for any first-order sentence $\phi$ we have
$$
\pi\sem{\phi}{} ~=~  \mathrm{dual}(\pi\sem{\neg\phi}{})
$$
\end{proposition}
Since $\pi\sem{\neg\neg\phi}{}=\pi\sem{\phi}{}$ it follows that also $\pi\sem{\neg\phi}{} ~=~  \mathrm{dual}(\pi\sem{\phi}{})$.

\subsection{Confidence Maximization} 
\label{subsec:conf-max}

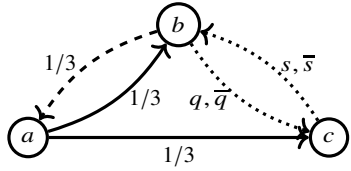
\begin{figure}
\centering
\begin{tikzpicture}
\begin{scope}
  \node[shape=circle,draw=black,very thick] (A) at (0,0) {$a$};
  \node[shape=circle,draw=black,very thick] (B) at (2,1.5) {$b$};
  \node[shape=circle,draw=black,very thick] (C) at (4,0) {$c$};

 \draw[->,very thick] 
      (A) edge[bend right=20] node[right] {$\,1/3$} (B);
 \draw[->,very thick,dotted] 
      (B) edge[bend right=20] node[left] {$q,\nnq\,$} (C);
 \draw[->,very thick] (A) edge node[below] {$1/3$} (C);
 \draw[->,very thick,dotted] 
      (C) edge[bend right=20] node[right] {$\,s,\nns$} (B);
 \draw[->,very thick,dashed] 
      (B) edge[bend right=20] node[left] {$1/3\,$} (A);
\end{scope}
\end{tikzpicture}
\vspace*{-3mm}
\caption{Maximum confidence model with dominant vertex}
\label{fig:max-conf}
\end{figure}

As in Sect.~\ref{subsec:conf} we use the Viterbi semiring $\Vit$
from Sect.~\ref{sec:commutative-semirings} interpreting its values as
confidence scores.  
Interestingly, we can reverse analyze the provenance polynomials 
and use confidence scores to find a model in which confidence is
maximized.

In the context of the example in Sect.~\ref{subsec:rev-ex},
suppose that we have confidence $1/3$ in all the literals
that the \ModComp\ interpretation $\pi$ maps to
a (positive or negative) provenance token. This yields 
a $\Vit$-interpretation $\pi'$ which, by Proposition~\ref{prop:hom}
and~Proposition~\ref{prop:prov-univ},
factors as $\pi'= h\circ\pi$  where 
$h$ is the unique semiring homomorphism
$\poly{\Nat}{X\cup\nnX}\rightarrow\Vit$ that maps all the tokens
$p,\ldots,\nnp,\ldots$ to $1/3$ (this is perfectly plausible, as 
confidence is \emph{not} probability).

Now recall from Sect.~\ref{subsec:rev-ex}
the sentence $\neg\phi$ (which asserts that there exists a dominant
vertex).  We have computed $\pi\sem{\neg\phi}{} ~=~ pr\nnt + \nnp
q\nns t$. Obviously, $\pi'$ is inconsistent so further applying
$h(pr\nnt + \nnp q\nns t)= 1/27+1/81= 4/81$ is not
meaningful. However, we know from Corollary~\ref{cor:model-check} that each
monomial in $\pi\sem{\neg\phi}{}$ corresponds to some model of $\neg\phi$. In
this case we have exactly two proof tree choices, corresponding to different
models, and they give different
confidence to $\neg\phi$. To maximize confidence we choose the
monomial $pr\nnt$ therefore a model in which we have an edge $Eab$,
an edge $Eac$ and \emph{no} edge $Eba$.  This will ensure the
  dominance of vertex $a$ with confidence $1/27$, in other words,
$\neg\phi$ is $1/27$-true in this model.  
This model is shown in Figure~\ref{fig:max-conf} (the
 edge $Eba$ is dashed because it is absent but we still wanted to show
  the confidence 1/3 in this absence). 
The edges $Ebc$ and $Ecb$ are dotted because neither their presence
nor their absence contradicts the provenance assumptions. We can, in fact, 
continue with a provenance analysis for these two edges if other properties
of the model are of interest.

\section{From Model Update to Provenance Update}
\label{sec:upd}

In this section we indicate a method for updating provenance polynomials
corresponding to a \ModDef\ interpretation when the associated model is updated by inserting or deleting facts, without recomputing the
provenance polynomial from scratch. One can think of this method
as \emph{incremental provenance maintenance}, and relate it to incremental view maintenance~\cite{BunemanC79}.

For example, recall from Sect.~\ref{subsec:ex} the interpretation $\beta$ and
the structure $G$ that it defines (Figure~\ref{fig:modelG}), and the sentence
$\phi$ asserting ``no dominant vertex''. We had computed
$$
\beta\sem{\phi}{} ~=~ (\nnr+t)\cdot p\cdot(1+q+\nns).
$$
First suppose that we update $G$ by \emph{deleting} $Eab$ and $Ebc$. 
Let us also assume that in the resulting model, we wish to track the presence/absence of 
the edges unaffected by the update, as well to introduce tokens that track the absence
of the two newly deleted edges. This results in the interpretation and the model
$\calH$ depicted in Figure~\ref{fig:modelH}. 
\begin{figure}[h]
\centering
\begin{tikzpicture}
\begin{scope}
  \node[shape=circle,draw=black,very thick] (A) at (0,0) {$a$};
  \node[shape=circle,draw=black,very thick] (B) at (2,1.5) {$b$};
  \node[shape=circle,draw=black,very thick] (C) at (4,0) {$c$};

 \draw[->,very thick,dashed] 
      (A) edge[bend right=20] node[right] {$\,\nnp$} (B);
 \draw[->,very thick,dashed] 
      (B) edge[bend right=20] node[left] {$\nnq\,$} (C);
 \draw[->,very thick,dashed] (A) edge node[below] {$\nnr$} (C);
 \draw[->,very thick,dashed] 
      (C) edge[bend right=20] node[right] {$\,\nns$} (B);
 \draw[->,very thick] 
      (B) edge[bend right=20] node[left] {$t\,$} (A);
\end{scope}
\end{tikzpicture}
\vspace*{-3mm}
\caption{The model $\calH$}
\label{fig:modelH}
\end{figure}

What is the 
corresponding update on the dual polynomial $\beta\sem{\phi}{}$? For the
provenance polynomials used for positive queries, as in~\cite{GreenKarTan07},
this update is performed by setting $p=q=0$. However, this would result
in the polynomial 0. Of course this cannot be right because $\calH\models\phi$. 
The deeper reason why it's wrong is that in $\beta$ we already had the literal $\neg Eab$ interpreted as 0. If we also interpret $Eab$ as $p=0$ the resulting interpretation does not define any model, much less $\calH$, and~Proposition~\ref{prop:all-proofs-mod-def} does not apply. The same issue arises with setting $q=0$.

Instead, the right way to perform this update takes advantage of the results 
in Sect.~\ref{subsec:prop-prov}. We use the \ModComp\ interpretation $\pi$
given in Sect.~\ref{subsec:rev-ex}.  Any other \ModComp\ interpretation
that both $G$ and $\calH$ are compatible with and that specializes
with respect to $G$ to $\beta$ would do, but $\pi$ is in an obvious sense the most
``economical'' such. Recall from Sect.~\ref{subsec:rev-ex} that
$$
\pi\sem{\phi}{} ~=~ 
(\nnp+\nnr+t)\cdot(p+\nnq+s+\nnt)\cdot(1+q+r+\nns),
$$
and therefore
$$
\refi{\pi}{\calH}\sem{\phi}{} ~=~ 
(\nnp+\nnr+t)\cdot\nnq\cdot(1+\nns)
$$
is the update we desire. 

Next, suppose that we update $G$ by \emph{inserting} $Eac$ and $Ecb$
resulting in the model $\calF$ in Figure~\ref{fig:modelF}. Then, the update
of $\beta\sem{\phi}{}$ is
$$
\refi{\pi}{\calF}\sem{\phi}{} ~=~ 
t\cdot(p+s)\cdot(1+q+r).
$$

To prove that this update method is sound,
we describe it more generally.
Let $\beta:\LitA(\tau)\ra\N[X,\nnX]$ be a model-defining interpretation which describes the model $\AA_\beta$ and 
tracks the facts 
that can potentially be added to or deleted from $\AA_\beta$.
More precisely, the set of these facts is $R^+\cup R^-$ 
where 
\begin{align*}  
R^+&=\{\alpha\in\FactsA(\tau): \beta(\alpha)=0 \text{ and } \beta(\neg\alpha)=\nnx_\alpha\} \quad\text{ and }\\
 R^-&=\{\alpha\in\FactsA(\tau): \beta(\alpha)=x_\alpha \text{ and } 
 \beta(\neg\alpha)=0\}.
\end{align*} 
An update can be any subset $U\subseteq R^+\cup R^-$ and it
updates the model $\AA_\beta$ to a new model $\AA_\beta[U]$,
with its associated model defining $\N[X,\nnX]$-interpretation
$\beta_U$.

Given any sentence $\phi\in\FO$ and its provenance polynomial
$\beta\ext\phi$ our method permits us to compute, for every
given update $U\subseteq R^+\cup R^-$, the updated provenance
polynomial $\beta_U\ext\phi$, without explicitly computing
the entire interpretation $\beta_U$.

For this purpose, we consider the general \emph{model-compatible}
interpretation $\pi$ with $\pi(\alpha)=x_\alpha$ and
$\pi(\neg\alpha)=\nnx_\alpha$ for every $\alpha\in R^+\cap R^-$
(and which maps all other literals to their truth values). This
interpretation is compatible with the class of all 
updated structures $\AA_\beta[U]$,
for $U\subseteq R^+\cap R^-$.
We now compute the provenance polynomial $\pi\ext\phi$ for
the model-compatible interpretation $\pi$ and use it
to derive the polynomials $\beta_U\ext\phi$ for
any given $U$.

\begin{proposition} For every update $U$ the provenance polynomial
$\beta_U\ext\phi$ can be computed from $\pi\ext\phi$ by setting
$x_\alpha=0$ for $\alpha\in ((R^+\setminus U) \cup (R^-\cap U))$
and $\nnx_\alpha=0$ for $\alpha\in ((R^+\cap U) \cup (R^-\setminus U))$.
\end{proposition}

The proof is an immediate consequence of the fact that 
the substitutions setting $x_\alpha, \nnx_\alpha$ to 0 as described in the proposition are precisely those specializing the 
model-compatible interpretation $\pi$ to the model defining interpretation $\beta_U$. Indeed they set to
0 precisely the indeterminates associated with the literals that
are either already false in $\AA_\beta$ and not changed by $U$
or true in $\AA_\beta$ but changed by $U$.

\section{Explanations and Repairs}
\label{sec:repairs}

In this section we discuss \emph{why-not} provenance questions. These arise as we wish to understand why a certain tuple does \emph{not} appear in the answer to a query, i.e., it is \emph{missing}~\cite{HerschelHT09,MeliouGMS10,HerschelH10} as well as when we wish to understand why 
\emph{integrity constraints} fail in a model. A first stab at the material in this section was made in~\cite{XuZhaAlaTan18}. For a very recent related paper see~\cite{BogaertsJV24}.

In databases, many integrity constraints are first-order expressible and \emph{repairs} for them have been studied in the context of query answering over inconsistent databases (i.e., databases in which integrity constraints fail (see~\cite{Bertossi19} and the many references in there). We will discuss both explanations and repairs for both missing answers and integrity constraint failure as problems that can be approached using dual provenance polynomials.

The approach proposed in this paper for dealing with negation in queries allows us to treat such questions in the same way in which the provenance polynomials in~\cite{GreenKarTan07} were used to answer questions about the presence of tuples in the answer and derive explanations for why-not questions in the same manner as in~Sect.~\ref{subsec:expl}.

We try to explain this method first via examples for missing query answers and failures of integrity constraints, and we will then formulate it in a more general way, as for updates of provenance polynomials in the previous section.

\subsection{Missing Query Answers}

\begin{figure}[ht]
\begin{center}
\begin{tikzpicture}
\begin{scope}
  \node[shape=circle,draw=black,very thick] (A) at (0,0) {$a$};
  \node[shape=circle,draw=black,very thick] (B) at (2,1.5) {$b$};
  \node[shape=circle,draw=black,very thick] (C) at (4,0) {$c$};

 \draw[->,very thick,dashed] 
     (A) edge[bend right=20] node[right] {$\nnp$} (B);
 \draw[->,very thick] 
      (B) edge[bend right=20] node[left] {$q$} (C);
 \draw[->,very thick] (A) edge node[below] {$r$} (C);
 \draw[->,very thick,dashed] 
      (C) edge[bend right=20] node[right] {$\nns$} (B);
 \draw[->,very thick] 
      (B) edge[bend right=20] node[left] {$t$} (A);
\end{scope}
\end{tikzpicture}
\end{center}
\caption{Model $\calM$}
\label{fig:modelM}
\end{figure}

Consider the model and the model-defining provenance-tracking interpretation in~Fig.~\ref{fig:modelM}. Let's call this
interpretation $\delta$.
Following the same visual conventions as with the model-defining, provenance-tracking interpretation $\beta$ illustrated in~Fig.~\ref{fig:modelG} we have:

\[ \delta(L) ~=~ \begin{cases}
           0  & \mathrm{if~} L=Eab\\
        \nnp  & \mathrm{if~} L=\neg Eab\\
           q  & \mathrm{if~} L=Ebc\\
           0  & \mathrm{if~} L=\neg Ebc\\
           r  & \mathrm{if~} L=Eac \\
           0  & \mathrm{if~} L=\neg Eac
            \end{cases}
\qquad=\qquad
~\begin{cases}
           0  & \mathrm{if~} L= Ecb\\
        \nns  & \mathrm{if~} L=\neg Ecb\\
           t  & \mathrm{if~} L= Eba\\
           0  & \mathrm{if~} L=\neg Eba\\
           0  & \mathrm{for~the~other~positive~facts}\\
           1  & \mathrm{for~the~other~negative~facts.}
            \end{cases}
\]

We also consider the query
$\domi(x) ~=~ \forall y\: (x=y \vee (Exy\wedge \neg Eyx))$.
$b$ is an answer for the query. Indeed, the provenance of ~$\domi(b)$~ 
under $\delta$ is~~$(0+t\nnp)(1+(0\cdot0))(0+q\nns) = \nnp q\nns t$.
Now suppose we ask:

\medskip\emph{Why is $a$ not an answer to this query?} 

\medskip
Using~Proposition~\ref{prop:all-proofs-mod-def} we reason as follows.
Since $\domi(a)$ is not valid in the model $\calM$, its provenance under $\delta$ must be the 
0 dual polynomial. However, its negation, namely (put in negation normal form) 
$$
\neg\domi(x) ~=~ \exists y (x\neq y \wedge (\neg Exy\vee Eyx)).
$$
is valid in $\calM$ and, as \emph{explanations} (causes) for the missing answer $a$, we will
take (see~Sect.~\ref{subsec:expl}) the reasons the provenance under $\delta$ of $\neg\domi(a)$ is \emph{not} the 0 polynomial.

Calculating this provenance for the interpretation $\delta$ gives
$$
0\cdot(0+1) + 1\cdot(\nnp+t) + 1\cdot(0+0) ~=~ \nnp+t
$$
This leads to two alternative explanations (causes) that answer $(\ast)$:
\begin{enumerate}
\item ~~$\nnp\neq0$~~ (absence of edge $Eab$)
\item ~~$t\neq0$~~ (presence of edge $Eba$)
\end{enumerate}

In addition to explanations, this methodology can also indicate \emph{repairs}, that is, updates to the model
$\calM$ that would produce a model that makes $\domi(a)$ true. This will be achieved by solving the equation
$$
\delta\sem{\neg\domi(a)}{} = 0
$$
This means $\nnp+t=0$ and therefore $\nnp=t=0$. We interpret $\nnp=0$ as ``insert $Eab$'' and $t=0$ as ``delete $Eba$''.

\val{Explain why this is sound, that is we compute an update which when applied to the model, transforms it into a model of
$\domi(a)$. Also why this is complete. Define complete?!? Either this is the only repair or any repair contains this repair.
but see the next example where we have multiple minimal repairs...}

\val{I am guessing that the above issues might be made more clear with a model-compatible (MULTI-MODEL) interpretation 
so here we go.}

Consider a model-compatible interpretation that uses the tokens used by $\delta$, namely $\nnp,q,r,\nns,t$. As a model-compatible interpretation, it must also use $p,\nnq,\nnr,s,\nnt$. The interpretation $\pi$ described in~Fig.~\ref{fig:assump} is such an interpretation. We have
\begin{eqnarray*}
\pi\sem{\domi(a)}{} & = & (1+0\cdot1)(0+p\cdot\nnt)(0+r\cdot1) ~=~ p\nnt r\\
\pi\sem{\neg\domi(a)}{} & = &  0\cdot(1+0) + 1\cdot(\nnp+t) + 1\cdot(\nnr+0) ~=~ \nnp+t+\nnr
\end{eqnarray*}
We could have started with this model-compatible interpretation and then we would specialize it (as in~Definition~\ref{def:spec}) to the model $\calM$ in~Fig.~\ref{fig:modelM} obtaining the $\delta$ above as $\delta = \refi{\pi}{\calM}$.

Now let's denote by $\calM^u$ the model $\calM$ updated with the repair we obtained: 
insert $Eab$ and delete $Eba$. 
As expected, $\refi{\pi}{{\calM^u}}\sem{\neg\domi(a)}{}=0$ and $\refi{\pi}{{\calM^u}}\sem{\domi(a)}{} = p\nnt r$ is the updated provenance
of $\domi(a)$.

\subsection{Integrity Constraint Failure}

Consider a slightly different model and interpretation.

\begin{figure}[ht]
\begin{center}
\begin{tikzpicture}
\begin{scope}
  \node[shape=circle,draw=black,very thick] (A) at (0,0) {$a$};
  \node[shape=circle,draw=black,very thick] (B) at (2,1.5) {$b$};
  \node[shape=circle,draw=black,very thick] (C) at (4,0) {$c$};

 \draw[->,very thick,dashed] 
     (A) edge[bend right=20] node[right] {$\nnp$} (B);
 \draw[->,very thick,dashed] 
      (B) edge[bend right=20] node[left] {$\nnq$} (C);
 \draw[->,very thick] (A) edge node[below] {$r$} (C);
 \draw[->,very thick] 
      (C) edge[bend right=20] node[right] {$s$} (B);
 \draw[->,very thick] 
      (B) edge[bend right=20] node[left] {$t$} (A);
\end{scope}
\end{tikzpicture}
\end{center}
\caption{Model $\calC$}
\label{fig:modelC}
\end{figure}

Consider also the integrity constraint (IC): ``\textsc{At least one vertex is dominant}''
$$
\exists x\,\domi(x)
~=~  \exists x\, \forall y\: (x=y \vee (Exy\wedge \neg Eyx))
$$

As before, we ask
\textit{``why is the IC failing in $\calC$?''}.

Note that the interpretation illustrated in~Fig.~\ref{fig:modelC} indicates the facts in $\calC$ that we are interested in tracking in explanation and willing to involve in repairs. The same facts and the tokens used to label them appear in the model-compatible interpretation $\pi$ that was used for the missing answers example so we can use it it again. We have (note that both these polynomials were calculated in~\ref{subsec:rev-ex})
\begin{eqnarray*}
\pi\sem{\exists x\,\domi(x)}{} & = & pr\nnt + \nnp q\nns t\\
\pi\sem{\neg\exists x\,\domi(x)}{} & = & (\nnp+\nnr+t)\cdot(p+\nnq+s+\nnt)\cdot(1+q+r+\nns)
\end{eqnarray*}

As expected $\refi{\pi}{\calC}\sem{\exists x\,\domi(x)}{} =0$.
We look for explanations in 
$$
\refi{\pi}{\calC}\sem{\neg\exists x\,\domi(x)}{} 
~=~ (\nnp+t)\cdot(\nnq+s)\cdot(1+r) ~=~ \nnp \nnq+\nnp s+t\nnq+ts + 
\nnp \nnq r+\nnp sr+t\nnq r+tsr
$$
This gives 8 alternative explanations but 4 of them are ``redundant''. We are left with 4 minimal explanations:
\val{Discuss redundant and minimal?}

\begin{enumerate} 
\item
 ~~$\nnp\nnq\neq0$~~ (absence of $Eab$ and $Ebc$)
\item
~~$\nnp s\neq0$~~ (absence of $Eab$ and presence of $Ecb)$
\item
~~$t\nnq\neq0$~~ (presence of $Eba$ and absence of $Ebc)$
\item
~~$ts\neq0$~~ (presence of $Eba$ and presence of $Ecb)$
\end{enumerate}

\eat{
For explanations, \emph{dualize} the tree:

\bigskip

\begin{center}
\begin{tikzpicture}
\begin{scope}
  \node[shape=circle,text width=5mm, draw=black,very thick] 
        (A) at (7.5,5) {$\!\!$and}; 
\node[shape=circle,text width=5mm, draw=black,very thick] 
        (B) at (1.5,2) {or};
\node[shape=circle,text width=5mm, draw=black,very thick] 
        (C) at (7.5,2) {or};
\node[shape=circle,text width=5mm, draw=black,very thick] 
        (D) at (13.5,2) {or};
  \node (E) at (0,0) {$\nnp\neq0$};
  \node (F) at (3,0) {$t\neq0$};
  \node (G) at (6,0) {$\nnq\neq0$};
  \node (H) at (9,0) {$s\neq0$};
  \node (I) at (12,0) {$1\neq0$};
  \node (J) at (15,0) {$r\neq0$};

\draw[-,very thick] (A) edge node[below] {} (B);
\draw[-,very thick] (A) edge node[below] {} (C);
\draw[-,very thick] (A) edge node[below] {} (D);
\draw[-,very thick] (B) edge node[below] {} (E);
\draw[-,very thick] (B) edge node[below] {} (F);
\draw[-,very thick] (C) edge node[below] {} (G);
\draw[-,very thick] (C) edge node[below] {} (H);
\draw[-,very thick] (D) edge node[below] {} (I);
\draw[-,very thick] (D) edge node[below] {} (J);
\end{scope}
\end{tikzpicture}
\end{center}

\bigskip


Four minimal \textbf{explanations}:\\

$~~\{\nnp\neq0,\nnq\neq0\}~~$
$~~\{\nnp\neq0,s\neq0\}~~$
$~~\{t\neq0,\nnq\neq0\}~~$
$~~\{t\neq0,s\neq0\}~~$
}

In order to determine repairs to the model $\calC$ that make the IC true we
solve the equation
$$
(\nnp+t)\cdot(\nnq+s)\cdot(1+r) ~=~ 0
$$
The minimal solutions appear in the following and-or tree:

\bigskip

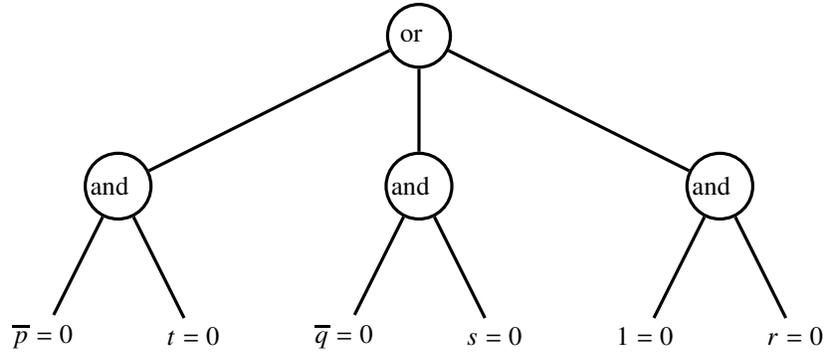
\begin{center}
\begin{tikzpicture}
\begin{scope}
  \node[shape=circle,text width=5mm, draw=black,very thick] 
        (A) at (5,4) {or}; 
\node[shape=circle,text width=5mm, draw=black,very thick] 
        (B) at (1,2) {$\!\!$and};
\node[shape=circle,text width=5mm, draw=black,very thick] 
        (C) at (5,2) {$\!\!$and};
\node[shape=circle,text width=5mm, draw=black,very thick] 
        (D) at (9,2) {$\!\!$and};
  \node (E) at (0,0) {$\nnp=0$};
  \node (F) at (2,0) {$t=0$};
  \node (G) at (4,0) {$\nnq=0$};
  \node (H) at (6,0) {$s=0$};
  \node (I) at (8,0) {$1=0$};
  \node (J) at (10,0) {$r=0$};

\draw[-,very thick] (A) edge node[below] {} (B);
\draw[-,very thick] (A) edge node[below] {} (C);
\draw[-,very thick] (A) edge node[below] {} (D);
\draw[-,very thick] (B) edge node[below] {} (E);
\draw[-,very thick] (B) edge node[below] {} (F);
\draw[-,very thick] (C) edge node[below] {} (G);
\draw[-,very thick] (C) edge node[below] {} (H);
\draw[-,very thick] (D) edge node[below] {} (I);
\draw[-,very thick] (D) edge node[below] {} (J);
\end{scope}
\end{tikzpicture}
\end{center}

Each solution corresponds to a different \emph{alternative repair}:
\begin{enumerate}
\item
$~~\{\nnp=t=0\}~~$ (insert $Eab$ and delete $Eba$), or
\item
$~~\{\nnq=s=0\}~~$ (insert $Ebc$ and delete $Ecb$)
\end{enumerate}

We can now update the provenance of $\exists x\,\domi(x)$
according to each of the repairs. Let $\calC^u_1$
be the model obtained by updating $\calC$ with the first repair. We obtain
$\refi{\pi}{{\calC^u_1}}\sem{\exists x\,\domi(x)}{} = pr\nnt$.
And let $\calC^u_2$
be the model obtained by updating $\calC$ with the second repair. We obtain
$\refi{\pi}{{\calC^u_2}}\sem{\exists\, x\domi(x)}{} = \nnp q\nns t$.

We can use this updated provenance to choose between the two repairs based on a provenance analysis, for example by cost (using the tropical semiring, $\Trop$). Toward this we make the following  \emph{assumptions:}  cost of one insertion: 20, cost of one deletion: 15; cost of pos/neg facts in the model $\calC$ initially, are
$\mathrm{cost}(\nnp)=\mathrm{cost}(\nnq)=10$,
$\mathrm{cost}(s)=\mathrm{cost}(t)=5$,
$\mathrm{cost}(r)=10$.

Then, $\mathrm{cost}(pr\nnt) = 20 + 10 + 15=45$
and
$\mathrm{cost}(\nnp q\nns t) = 10 + 20 + 15 + 5=50$. We conclude that the first repair is cheaper.

\val{What follows are some facts about the complexity of this whole method, some I know how to prove , some not.}

In general, there can be an exponential number of minimal repairs. When we calculate the dual polynomial associated to a sentence we should leave it 
in expression form rather than put it in polynomial form because the latter may have exponential size. The expression can stay in polysize. When we solve the equation for repairs, the and-or tree of repairs is also of polynomial size.
(All this is data complexity.)
Any minimal repair is a subset of a repair represented
in the tree.

\subsection{Repairs by provenance polynomials}

In a general sense, the method to compute (minimal) repairs
is just a variation of the method computing updated provenance polynomials described in Sect.~\ref{sec:upd}.
Assume that $\AA$ is a $\tau$-structure with universe A such that $\AA\not\models\psi$. 
Assume further that to repair $\AA$
and make it a model of $\psi$, we have sets $R^-$ and $R^+$
of facts that we are allowed to remove from or add to $\AA$.

As in Sect.~\ref{sec:upd} we have a provenance-tracking model-defining   interpretation $\beta:\LitA(\tau)\ra\N[X,\nnX]$ 
which defines the model $\AA_\beta=\AA$ and 
tracks the facts in $R^+\cup R^-$
that can potentially be added to or deleted from $\AA$.

A \emph{repair} for $\AA$ and $\psi$
is now an update $R\subseteq R^+\cup R^-$ 
such that the modified structure $\AA[R]$ is a model for $\psi$.

Let $X$ be the set of indeterminates $x_\alpha$ where $\alpha\in R^+\cup R^-$. We again consider the \emph{model-compatible} interpretation $\pi:\Lit_A(\tau)\ra\N[X,\nnX]$
such that, for every atom $\alpha\in R^+\cup R^-$, we have 
$\pi(\alpha)=x_\alpha$ and $\pi(\neg\alpha)=\bar{x_\alpha}$,
and all other literals
$\beta=\alpha$ or $\beta=\neg\alpha$
where $\alpha$ is \emph{not} in $R^+\cup R^-$ (i.e. can not be changed)
is simply mapped it to its truth value in $\AA$. 

We now consider the provenance polynomial $\pi\ext\psi\in\N[X,\bar X]$ and the subset 
$X^\pm = \{ x_\alpha : \alpha \in R^+ \} \cup 
\{ \bar{x_\alpha} : \alpha \in R^- \} \subseteq X\cup\nnX$
of the dual indeterminates that are directly associated with the possible changes.

We write $\pi\ext\psi=m_1+\dots +m_k$
as a sum of monomials, and let $v(m)$ be the
set of variables from $X\cup \nnX$ appearing in $m$.
However, not all these variables are necessarily in 
$X^{\pm}$, since $X\cup\nnX$ also contains their duals.
By examining what combinations of indeterminates from $X^\pm$ 
occur in the monomials of $\pi\ext\psi$, 
we can read off all minimal repairs.

\begin{proposition}\label{PropRepairs}
For every monomial $m$ of  $\pi\ext\psi$ the set
\[ R_m:=\{\alpha\in R^{\pm}: x_\alpha\in v(m)\cap X^{\pm} \text{ or }
 \nnx_\alpha\in v(m)\cap X^{\pm}\}\]
is a repair of $\AA$ for $\psi$.
Conversely, for every repair $R \subseteq R^+\cup R^-$ of $\AA$ for $\psi$,
there is a monomial $m$ of $\pi \ext\psi$ such that
$R_m \subseteq R$. If $R$ is a minimal repair, then $R_m = R$.
\end{proposition}

\begin{proof}
The first claim follows directly from Theorem~\ref{thm:modelcompatible}.
Indeed, every monomial $m$  in $\pi\ext\psi$ corresponds to a proof
tree for $\psi$ that uses only literals that are true in $\AA[R_m]$
(because such a literal is either already true in $\AA$ or
is changed to true by $R_m$) 
so $\AA[R_m]\models\psi$. Suppose now that $R$ is any repair of $\AA$
for $\psi$. As $\AA[R]$ differs from $\AA$ only for literals 
in $R^{\pm}$  there is a unique assignment $h \colon X \cup \nnX \to \Bool$ such that $h \circ \pi$ is the $\Bool$-interpretation that
describes $\AA[R]$, and therefore 
$h \circ \pi\ext\psi  = h( \pi\ext\psi)=1$.
So there must be a monomial $m \in \pi\ext\psi$ with 
$h(m) = 1$.
For every variable $x_\alpha\in v(m)\cap X^{\pm}$ we have that
$h(x_\alpha)=1$ and hence $\alpha\in R$. For $\nnx_\alpha\in v(m)\cap X^{\pm}$
we have $h(\nnx_\alpha) = 1$ and thus again 
$\alpha \in R$ by construction of $h$.
This proves that $R_m \subseteq R$.
If $R$ is minimal, we have equality, because otherwise $R_m$
would be a smaller repair.
\end{proof}

We remark that instead of $\N[X,\nnX]$ one can use as well simpler semirings
such as $\Sorb(X,\nnX)$ or even $\PosBool(X,\nnX)$, for which the
provenance polynomial $\pi\ext\psi$ is smaller and easier to compute. Indeed,
even the polynomial in $\PosBool(X,\nnX)$ suffices to determine the minimal repairs,
since these only depend on the sets of variables occurring in some monomial.
However, a computation in a more informative semiring may have the advantage that we can compare the different repairs according to costs, clearance levels, 
confidence scores etc., for which the $\PosBool$-semirings are not sufficient.

\medskip An alternative road towards the computation of minimal repairs is based on computing not the provenance polynomial $\pi\ext\psi$
but, as in the examples discussed above, the provenance polynomial
$\pi\ext{\neg\psi}$, to explain why $\AA\models\neg\psi$.

Consider the two provenance polynomials $\pi\ext\psi$ and $\pi\ext{\neg \pi}$ and specialise them to $\AA$ by setting
$x_\alpha$ to 0, if $\AA\models\neg\alpha$ and $\nnx_\alpha=0$
if $\AA\models\alpha$.
For the resulting model-defining interpretation $\refi{\pi}{\AA}$
we obviously have that $\refi{\pi}{\AA}\ext{\psi}=0$ (since $\AA\not\models\psi$).

To compute repairs, we then compute solutions for the
equation $\refi{\pi}{\AA}\ext{\neg\psi}=0$.
A minimal solution is a minimal set of indeterminates
such that setting these to 0 annihilates 
$\refi{\pi}{\AA}\ext{\neg\psi}$.
If $\pi\ext\phi$, and hence also $\refi{\pi}{\AA}$, are written as a sum of monomials, we can write $\refi{\pi}{\AA}\ext{\neg\psi}$
as a product of sums of indeterminates, and 
compute solutions by setting each of these sums individually to 0.
When we have computed such a set $Y$, we get
the repair 
\[ R_Y:=\{\alpha \in R^+: \nnx_\alpha\in Y\} \cup 
\{\alpha \in R^-: x_\alpha\in Y\}.\]
Indeed, for $\BB=\AA[R_Y]$ the model-defining interpretation $\refi{\pi}{\BB}$
is obtained from $\pi$ by setting to 0 the indeterminates in $Y$ 
as well as the variables associated with literals that are false in
$\AA$ and are not touched by $R_Y$. 
It follows that
$\refi{\pi}{\BB}\ext{\neg\phi}=0$.
But since $\refi{\pi}{\BB}$ is model-defining this implies that
$\refi{\pi}{\BB}\ext{\phi}\neq 0$ and hence that $\BB\models\phi$.

\section{Other Approaches to Modeling Negation}

The semiring semantics for first-order logic outlined in this paper handles negation via transformation to negation normal form and polynomials with dual indeterminates. This has meanwhile emerged as the standard approach for dealing with negation, not just in first-order logic, but also many other logical systems.
One of the issues with this approach (but we actually consider it as a feature!) is that negation is not a compositional algebraic operation, contrary to disjunction and conjunction. Indeed, a semiring valuation of $\neg\phi$ is, in general, not uniquely determined by the valuation of $\phi$, but depends on the \emph{syntax} of $\phi$. Also, equivalences are not always preserved under negation, and 
the usual logical rules involving negation do not necessarily hold for all semirings.
Indeed, consider a semiring $\Semi$ where addition is idempotent, but multiplication is not, such as the tropical semiring. Then $\phi\lor\phi \equiv_\Semi \phi$, but in general 
$\pi\ext{\neg(\phi\lor\phi)}=\pi\ext{\nnf(\neg\phi)\land\nnf(\neg\phi)}=\pi\ext{\nnf(\neg\phi)}\cdot\pi\ext{\nnf(\neg\phi)}$ which is, in general, not the same as $\pi\ext{\neg\phi}=\pi\ext{\nnf(\neg\phi)}$.

However there are several alternative approaches to negation that we want to discuss here.

\subsection{Flattening} For any compositional interpretation of (non-atomic) negation
in a semiring $\Semi$, we need a function $f:\Semi\ra\Semi$ so that,
for every $\Semi$-interpretation $\pi$ and every (non-atomic) formula $\phi$, we can 
define the valuation $\pi\ext{\neg\phi}:=f(\pi\ext{\phi})$. If we wish to remain consistent with the intuition that 0 stands for \emph{false} and all other values 
$s\in\Semi$ correspond to an annotated variant of \emph{true} then $f$ must have the 
property that $f(s)=0$ for $s\neq 0$ and $f(0)=t$ for some fixed value 
$t\in\Semi$. The most reasonable choice is to take $t=1$.

Interpreting negation in this way has the consequence that, under the scope of negation,
semiring sematics reduces to Boolean semantics. For every non-atomic formula $\phi$
and every model-defining $\Semi$-interpretation $\pi$ we have that $\pi\ext{\neg\phi}=1$
if $\AA_\pi\models \neg\phi$ and $\pi\ext{\neg\phi}=0$
if $\AA_\pi\models \phi$. Further although in general $\neg\neg\psi\not\equiv_\Semi\psi$
for atomic formulae $\psi$, we obviously have that $\neg\neg\neg\psi\equiv_\Semi\neg\psi$
for every formula $\psi$.  We can conclude, that while the use of a flattening function $f$ permits to define a compositional negation operator that avoids the
transformation to negation normal form, this does not lead to interesting 
provenance information once we are under the scope of a negation.

\medskip
In special cases, for instance for semirings over the real interval $[0,1]$ such as the fuzzy semiring or the {\L}ukasiewicz semiring, and \emph{if we deviate from the standard intuition}, also other functions can make sense, such as $f(x):=1-x$.
We shall come back to this in the context of dealing with implication.

\medskip A more general approach towards negation is based on the observation that, classically, negation can be obtained from
other standard logical operations such as implication or difference. So we can try to define appropriate semiring semantics for
first-order logic with implication and/or difference, and then identify $\neg\psi$ with $\psi\ra 0$ or with $1-\psi$.

\subsection{Monus semirings}  

Monus is an operation on certain naturally ordered commutative monoids generalising the
special cases of the natural numbers, where it is subtraction, truncated to 0, and of the Boolean truth values
where it $x\land\neg y$.
Monus was equationally axiomatized by Bosbach \cite{Bosbach65} and Amer \cite{Amer84},
and it was introduced to semirings for provenance by Geerts and Poggi \cite{GeertsPog10}. 

Recall that we generally restrict attention to naturally ordered semirings $(S,+,\cdot,0,1)$, 
where $s\leq t :\leftrightarrow \exists r (s+r=t)$ is antisymmetric and hence defines a partial
order. This condition, and also the monus operation, only depend on 
the additive monoid $(S,+,0)$ of the semiring.
Natural order does not admit an equational characterization. It is therefore remarkable that with the addition of monus, the resulting structures form an algebraic variety.

\begin{proposition}
Let $(S,+,0)$ be a commutative monoid. For a binary operation
$a\dotdiv b$ on $S$ the following characterizations are equivalent
\begin{enumerate}
\item $S$ is naturally ordered and for any $a,b\in S$, $a\dotdiv b=\min\{c : a\leq b+c\}$.
\item $S$ is naturally ordered and for any $a,b,c\in S$, $a\dotdiv b\leq c$ if, and only if, $a\leq b+c$. \val{For any $b$, $\lambda a.a\dotdiv b$ and $\lambda c.b+c$ form a Galois connection.}
\item $S$ is naturally ordered and for any $b,c\in S$, $b+c$ is the largest $a$ such that $a\dotdiv b\leq c$.
\item The following four equational axioms hold
    \begin{enumerate}
    \item $a\dotdiv a = 0$
    \item $0\dotdiv a = 0$
    \item $a+(b\dotdiv a) = b+(a\dotdiv b)$
    \item $a\dotdiv(b+c) = (a\dotdiv b)\dotdiv c$
    \end{enumerate}
\end{enumerate}
\end{proposition}

\eat{
\val{SOME PROOFS I don't think we need to put this in but I want to record it somewhere}

(1),(2), and (3) are equivalent by general facts about Galois connections.

(4a) and (4b) follow immediately from (1). 
\val{In this direction, it remains to show that (1) also implies
proving (4c) and (4d)}

Note that (2) implies $a\leq b+(a\dotdiv b)$ and $a\dotdiv b\leq a$.
\val{Maybe useful.}

\bigskip
=============
\bigskip

}

Any idempotent commutative monoid is naturally ordered (in fact, the natural order is the semilattice order).

\begin{proposition}
Let $(S,+,0)$ be an idempotent commutative monoid. For a binary operation
$a\dotdiv b$ on $S$ the following characterizations are equivalent
\begin{enumerate}
\item For any $a,b\in S$, $a\dotdiv b=\min\{c : a\leq b+c\}$.
\item For any $a,b,c\in S$, $a\dotdiv b\leq c$ if, and only if, $a\leq b+c$. \val{For any $b$, $\lambda a.a\dotdiv b$ and $\lambda c.b+c$ form a Galois connection.}
\item For any $b,c\in S$, $b+c$ is the largest $a$ such that $a\dotdiv b\leq c$.
\item The following four equational axioms hold
    \begin{enumerate}
    \item $a\dotdiv a = 0$
    \item $(a\dotdiv b)+b = a+b$
    \item $(a\dotdiv b)+a=a$
    \item $(a+b)\dotdiv c = 
            (a\dotdiv c)+(b\dotdiv c)$
    \end{enumerate}
\end{enumerate}
\end{proposition}

\eat{
\val{SOME PROOFS}
Proving (5b), (5c) and (5d) from (2) when $+$ is idempotent \val{seems much easier!}. 

Indeed, (5c) follows from $a\dotdiv b\leq a$. For (5b) observe that it also follows that $(a\dotdiv b)+b\leq a+b$ and that from $a\leq b+(a\dotdiv b)$ 
it follows that $a+b\leq b+(a\dotdiv b)+b = b+(a\dotdiv b)$.

Finally, for (5d), by (2) we could show that 
$$
(a\dotdiv c)+(b\dotdiv c) = \min\{x\mid a+b\leq c+x\}
$$
Indeed, $a\leq c+(a\dotdiv c)$ and  $b\leq c+(b\dotdiv c)$
hence $a+b\leq 2c+(a\dotdiv c)+(b\dotdiv c)$ but $2c=c$.
Moreover, if $a+b\leq c+x$ for some $x$ then $a\leq c+x$
and $b\leq c+x$ hence $a\dotdiv c\leq x$ and $b\dotdiv c\leq x$
hence $(a\dotdiv c)+(b\dotdiv c)\leq x+x=x$.

Let's also \val{try to} prove, under the assumption that $+$ is idempotent, that
(5) implies (2).

First we show that $a\dotdiv b$ satisfies $a\leq b+(a\dotdiv b)$, that is, $a+b+(a\dotdiv b) = b+(a\dotdiv b)$. This follows from (5b) and (5c).

Next we show that if for some $x$ we have $a\leq b+x$ then $a\dotdiv b\leq x$. \val{Stuck. I presume we would somehow use also (5a) and (5d)}

\val{Note also that (4c) follows directly from (5b)}

\val{Note also that (4c) implies that $a\leq b+(a\dotdiv b)$.END OF PROOFS }
}

The following naturally ordered commutative monoids have a monus operation (unique by characterization 1):
\begin{itemize}
    \item $(\Nat,+,0)$, where monus is truncated subtraction, i.e. $a\dotdiv b=0$ if $a\leq b$. 
    \item $(B,\lor,\bot)$~~ where $(B,\lor,\land,\bot,\top,\neg)$ is a Boolean alegebra, and where monus is $a\land\neg b$. For the Boolean algebra of subsets of a set, $(2^X,\cup,\cap,\emptyset,X)$, monus is set difference.
    \item Let $(T,\leq)$ be a totally ordered set with least element (denoted $\bot$). Then $(T,\max,\bot)$ is a commutative monoid that is idempotent, hence naturally ordered, and that has monus:
    $$
    a\dotdiv b = \begin{cases}
                  \bot & \mbox{when}~a\leq b\\
                  a    & \mbox{when}~a> b\\
                 \end{cases}
    $$
    So we can define monus in the security semiring, the Viterbi semiring (hence the tropical semiring), the {\L}ukasiewicz semiring, and the fuzzy semiring \val{others?}
    \item Let $(L,\leq)$ be a complete lattice. $L$ has a least element (notation $\bot$) and binary join (notation $\sqcup$)
    so that $(L,\sqcup,\bot)$ is an idempotent commutative monoid. When $L$ is distributive
    \val{I use the distributivity of $\sqcup$ over $\inf$.
    I don't know if this condition is necessary.}
    this monoid has monus, by the following proposition.
\end{itemize}
    
\begin{proposition} Let $(L,\leq)$ be a distributive complete lattice. For any $a,b\in L$ let $z=\inf\{c\mid a\leq b\sqcup c\}$. Then
$a\leq b\sqcup z$.
\end{proposition}

Indeed, Let $C=\{c\mid a\leq b\sqcup c\}$. Then $b\sqcup z=
b\sqcup\inf C = \inf(b\sqcup C)$ and for any $x\in b\sqcup C$
we have $x=b\sqcup c$ for some $c\in C$ so $a\leq x$.

In particular for the distributive complete lattice $(2^X,\subseteq)$ the commutative monoid $(2^X,\cup,\emptyset)$
has monus. It better be the same as the one that emerges from the Boolean algebra structure! Indeed, one can check that for any $A,B\in2^X$
$$
\bigcap_{A\subseteq B\cup C} C ~=~ A\setminus B
$$

Moreover, any finite distributive lattice is also a complete distributive lattice, and thus this class of semirings, in particular $\PosBool(X)$, has monus. \val{We should calculate examples of monus in PosBool.}

\medskip

The monus operation gives a semiring interpretation of the difference operator of relational algebra or, generally for
``logical difference'': for any two formulae $\psi,\phi$, we can also admit the difference $\psi - \phi$ (which in Boolean semantics is the same as $\psi\land \neg\phi$)  and
define $\pi\ext{\psi - \phi}:=\pi\ext\psi\dotdiv\pi\ext\phi$. In absorptive semirings, where we have 1 as the maximal value, we can then
identify $\neg\psi$ with $1-\psi$ and define $\pi\ext{\neg\psi}:=1\dotdiv\pi\ext\psi$.
But does this give a ``reasonable'' interpretation for negation? Recall that on semirings
where $a+b=\max(a,b)$, we have that $a\dotdiv b=a$ whenever $a>b$ and $a\dotdiv b=0$ otherwise. Hence $\pi\ext{\neg\psi}=1$ whenever $\pi\ext\psi<1$, and $\pi\ext{\neg\psi}=0$
only in case $\pi\ext{\psi}=1$. Hence this ``negation'' is purely Boolean, and
is just the negation of ``$\psi$ has the maximal truth value"; in particular this is not consistent with the standard intuition that only 0 stands for \emph{false}, and all other values for an annotated variant of \emph{true}.
Further, the monus operator, and hence also negation, only depends on the additive monoid of the semiring, and does not take into account the interpretation of multiplication at all. Finally,
the correspondence of the difference $\psi-\phi$ with $\psi\land\neg\phi$ 
also breaks down on certain semirings. Consider for instance min-cost computations
in the tropical semiring $\Trop = (\R^\infty_+ , \min, +, \infty, 0)$ and assume that
$\psi$ and $\phi$ have the same cost $\pi\ext\psi=\pi\ext\phi$ which is neither 0 nor $\infty$.
Then the cost of $\psi-\phi$ is $\infty$ whereas the cost of $\psi\land\neg\phi$
is $\pi\ext{\psi}+\pi\ext{\neg\phi}=\psi\ext\psi + 0=\pi\ext{\psi}$.

\medskip
For all these reasons we conclude that while monus semirings provide an interpretation of (relational) difference that may be interesting in certain settings (such as for bag semantics on the natural semiring), the monus operation does not lead to a convincing valuation
for negation.

\subsection{Negation via implication}

A further possibility to deal with negation is based on the idea to define
an appropriate interpretation for implications $\psi\ra\phi$ and to
identify then $\neg\psi$ with $\psi\ra \bot$. But what is an appropriate
interpretation of $\psi\ra\phi$ and what properties of negation does it imply?
Implication corresponds to the semantic notion that $\psi$ \emph{entails} $\phi$, and should satisfy classcial properties such as modus ponens and the deduction theorem.
All this works best in the context of absorptive semirings, preferably   with additional properties that permit to define natural infinitary addition and multiplication operations, based on
infima and suprema \cite{BrinkeGraMrkNaa24}, so that we have natural semiring valuations also for infinite sets of formulae.

\begin{definition}\label{def:infinitary absorptive semirings}
An \emph{infinitary absorptive semiring} is based on an absorptive semiring $\Semi$ which satisfies the additional properties that
\begin{itemize} 
\item the natural order $(\calS, \le)$ is a complete lattice.
\item $\Semi$ is (fully) continuous:  for every non-empty chain $C\subseteq S$, the supremum $\Sup C$ and the infimum $\Inf C$ are compatible with
addition and multiplication, i.e.
\[ s\circ \Sup C = \Sup (s\circ C)  \quad\text{ and }  \quad s\circ \Inf C=\Inf  (s\circ C), \]
where $(s\circ C):=\{ s \circ  c : c\in C\}$ for every $s\in S$ and $\circ\in\{+,\cdot\}$.
\end{itemize}

As a consequence, we can define natural infinitary addition and multiplication operations in $\Semi$ (and thus
semiring provenance for first-order logic also on infinite universes) by taking suprema of finite subsums and infima of finite subproducts: 
\[  \sum_{i\in I} s_i :=\Sup_{\substack{ I_0 \subseteq  I\\I_0 \text{ finite}}}\Bigl( \sum_{i\in I_0} s_i\Bigr)\quad\text{ and }\quad  \prod_{i\in I} s_i :=\Inf_{\substack{ I_0 \subseteq  I\\I_0 \text{ finite}}}\Bigl( \prod_{i\in I_0} s_i\Bigr).\]
\end{definition}

Since addition is idempotent in absorptive semirings, the infinitary addition is in fact the same as the supremum:   $\sum_{i\in I} s_i = \Sup_{i\in I} s_i$.
However, unless multiplication is also idempotent (so that the semiring is a lattice semiring), infinitary products need not coincide with 
infima.

Let  $\pi:\Lit_A(\tau)\ra \Semi$ be a model-defining $\Semi$-interpretation for
a finite or infinite universe $A$ and a relational vocabulary $\tau$.
For sets of sentences $\Phi\subseteq\FO$, we put 
$\pi\ext{\Phi}:=\prod_{\phi\in\Phi} \pi\ext{\phi}$.

\begin{definition} Let $\Phi\subseteq\FO$ and $\psi\in\FO$, and let $\Semi$ be an infinitary absorptive semiring. We write
\begin{enumerate}
\item $\Phi\equiv_\Semi 0$ if $\pi\ext{\Phi}=0$ for every model-defining $\Semi$-interpretation $\pi$;
\item $\Phi\models_\Semi \psi$ if  $\pi\ext{\Phi}\leq \pi\ext{\psi}$  for every model-defining $\Semi$-interpretation $\pi$.
\end{enumerate}
\end{definition}

Notice that once we have committed ourselves to a specific semiring
$\Semi$, semiring semantics over $\Semi$ is a particular case of a multivalued logic,
and we may look for connections to interpretations of implication and negation in that area. 

\medskip\noindent{\bf The Gödel implication. }
We now add to an absorptive semiring  $\Semi$ the new binary operation $\ra$ mapping every pair $(s,t)$ to
\[  s\ra t:=\begin{cases} 1 &\text{ if }s\leq t\\t&\text{ otherwise.}\end{cases}\]
Further we use the abbreviation $\sim s:=s\ra 0$ which map 0 to 1 and all other elements of $\Semi$ to 0.

We can now extend propositional logic $\PL$ and first-order logic $\FO$ with semiring semantics
to $\PL^\ra$ and $\FO^\ra$ which permit to build formulae $\psi\ra\phi$ such that, for every 
$\Semi$-interpretation $\pi$ we have that
\[ \pi\ext{\psi\ra\phi}:=\pi\ext{\psi}\ra\pi\ext{\phi}=\begin{cases} 1 &\text{ if }\psi\entails\phi  \\ \pi\ext{\phi}&\text{ otherwise.}\end{cases}\]

Of course there are other possibilities to interpret the connective $\ra$ and thus other ways to extend the semiring semantics
of $\PL$ and $\FO$ by an interpretation for implication. We argue that the one we discuss here, which has originally been proposed
by Gödel,  has specifically interesting properties.
If the underlying semiring $\Semi$ is a min-max semiring then this interpretation of $\ra$ makes $\FO^\ra$
a Gödel logic. 

\begin{definition} Let $\Semi$ be an infinitary absorptive semiring with implication. We say that 
the (semantic)  deduction theorem (DT) holds for $\Semi$, i.e. if for all $\Phi\subseteq\FO$ and $\psi,\phi\in\FO$
\[ \Phi,\psi\models_\Semi \phi\quad\Iff\quad\Phi\models_\Semi \psi\ra\phi.\]
\end{definition}

Note that (DT) implies modus ponens: $\psi, \psi\ra\phi\models_\Semi\phi$. Informally, the deduction theorem expresses that the semantics of $\ra$ captures entailment in a sound and complete way.
For $\Phi=\emptyset$ this is trivially true and just says that an implication $\psi\ra\phi$ is
a strong tautology in $\Semi$ (i.e. evaluates to 1 under all $\Semi$-interpretations) if, and only if, $\psi\models_\Semi\phi$.

But under a non-empty set $\Phi$ of hypothesis, the situation is more complicated.
The soundness part of (DT), i.e.  $\Phi\models_\Semi \psi\ra\phi\ \Rightarrow\ \Phi,\psi\models_\Semi \phi$ holds for every infinitary absorptive semiring.
Indeed, if $\pi\ext{\psi}\leq\pi\ext{\phi}$ then also $\pi\ext{\Phi,\psi}\leq \pi\ext{\phi}$ and otherwise 
$\pi\ext{\psi\ra\phi}=\pi\ext{\phi}$ so $\Phi\models_\Semi \psi\ra\phi$ implies that 
$\pi\ext{\Phi,\psi}\leq\pi\ext{\Phi}\leq\pi\ext{\phi}=\pi\ext{\psi\ra\phi}$.
However, the completeness part of (DT) is only true for min-max semirings.

\begin{proposition}\label{DT-minmax} The deduction theorem holds for an infinitary absorptive semiring $\Semi$ if, and only if, $\Semi$ is a min-max semiring, i.e.
if, and only if, $\FO^\ra$ with semiring semantics given by $\Semi$ is a Gödel logic.
\end{proposition}
  
 \begin{proof} It is well-known that (DT) holds for Gödel logics. To see this it only remains to show that $\Phi,\psi\models_\Semi \phi\ \Rightarrow\ \Phi\models_\Semi \psi\ra\phi$.
 Assume that the left side holds and consider any $\Semi$-interpretation $\pi$. If $\pi\ext{\psi}\leq\psi\ext{\phi}$ then $\pi\ext{\psi\ra\phi}=1$ so
 $\pi\ext{\Phi}\leq \pi\ext{\psi\ra\phi}$ holds trivially. Otherwise $\pi\ext{\psi}>\psi\ext{\phi}$ so $\pi\ext{\Phi,\psi}=\min(\pi\ext{\Phi},\pi\ext{\psi})\leq\phi\ext{\phi}$
 implies that $\pi\ext{\Phi}\leq\pi\ext{\phi}=\pi\ext{\psi\ra\phi}$, so $\Phi\models_\Semi \psi\ra\phi$.
 
 To prove the converse it suffices to prove that (DT) fails if $\Semi$ is not multiplicatively idempotent or if $\Semi$ is not linearly ordered by the natural order.
 Indeed an absorptive and multiplicatively idempotent semiring is necessarily a lattice semiring and if the order is linear, it is a min-max semiring.   
 We first show that (DT) fails, even for propositional formulae, if the natural order on $\Semi$ is not a linear order, i.e. if there
exist incomparable element $s,t\in\Semi$.   For propositional variables $x,y$, let
$\Phi:=\{x\},\psi:=y$ and $\phi:=(x\land y)$. Obviously  $\Phi,\psi\models_\Semi \phi$. However, for the $\Semi$-interpretation
$\pi:x\mapsto s, y\mapsto t$ we have that $\pi\ext{\Phi}=\pi(x)=s$, but since $\pi\ext{\psi}=s\not\leq st=\pi\ext{\phi}$,
we have that $\pi\ext{\psi\ra\phi}=\pi\ext\phi=st$, so
$\Phi\not\models_\Semi \psi\ra\phi$. 
Assume next that $\Semi$ is not fully idempotent, i.e $s\cdot s < s$ for some $s\in\Semi$. Again it suffices to argue propositionally, this time
with a single variable $x$. Let $\Phi=\{x\}$, $\psi:=x$ and $\phi=(x\land x)$. Obviously  $\Phi,\psi\models_\Semi \phi$, but
setting $\pi(x):=s$, we have $\pi\ext{\Phi}=\pi\ext{\psi}=s$, but since $s> s\cdot s=\pi\ext{\phi}$,
we have that $\pi\ext{\psi\ra\phi}=\pi\ext\phi=s\cdot s$, so
$\Phi\not\models_\Semi \psi\ra\phi$. 
 \end{proof}

\medskip\noindent\textbf{A more general definition for implication. } 
 Proposition~\ref{DT-minmax}  raises the question, whether there are alternative interpretations for $\ra$ that satisfy (DT) on more general classes
 of semirings. We first observe that the Gödel implication is the only interpretation of $\ra$ on min-max semirings
 that is sound for entailment and satisfies (DT).

 \begin{proposition} Let $\Semi$ be a min-max semiring, and let $s\multimap t$ be an operation on $\Semi$ such that the induced 
 interpretation of $\psi\multimap \phi$ satisfies the following properties:
 \begin{itemize}
 \item If $\pi\ext{\psi}\leq\pi\ext{\phi}$, then $\pi\ext{\psi\multimap\phi}=1$;
 \item {\rm (DT)} holds for $\multimap$ on $\Semi$, i.e. $\Phi,\psi\models_\Semi \phi\ \Iff\ \Phi\models_\Semi \psi\multimap\phi$.
 \end{itemize}
 Then $\multimap$ is the Gödel implication, i.e. $(s\multimap t)=(s\ra t)$.
 \end{proposition}
 
\begin{proof} If $\pi\ext{\psi}\leq\pi\ext\phi$ then $\pi\ext{\psi\multimap\phi}=\pi\ext{\psi\ra\phi}=1$.
So assume that $\pi\ext{\psi}>\pi\ext{\phi}$. We have to prove that $\pi\ext{\psi\multimap\phi}=\pi\ext{\phi}$.
Since $\phi,\psi\models_\Semi \phi$ it follows by (DT) that $\phi\models_\Semi \psi\multimap\phi$, hence $\pi\ext{\phi}\leq \pi\ext{\psi\multimap\phi}$.
From $\psi\multimap\phi \models_\Semi \psi\multimap\phi$ we get by (DT) that $\psi\multimap\phi,\psi\models_\Semi \phi$.
Hence  $\min(\pi\ext{\psi\multimap\phi}\pi\ext{\psi})\leq\pi\ext\phi$. But since $\pi\ext{\psi}>\phi\ext{\phi}$ it follows that 
$\pi\ext{\psi\multimap\phi}\leq \pi\ext{\phi}$.
\end{proof} 
 
 We thus look for an more general interpretation of implication, which coincides with the Gödel implication on min-max semirings, but
 not necessarily on other semirings. What could be reasonable values $r$ for $s\ra t$?
 Since from $\psi$ and $\psi\ra\phi$ we want to able to infer $\phi$, we should have that $s\cdot(s\ra t)\leq t$.
 On the other side, the larger we define the value of $s\ra t$ the more powerful the reasoning with implications becomes.
 Thus, the following definition might be reasonable. Set 
 \[  s\ra t:=\sup\{ r:  r\cdot s\leq t\}\]

\begin{itemize}
 \item On min-max semirings this coincides with the Gödel implication.
 \item On the tropical semiring  $\Trop= (\mathbb{R}_{+}^{\infty},\min,+,\infty,0)$, we have that $s\ra t=\min(0, t-s)$
 (recall that the natural order on $\Trop$ is the inverse of the usual order on the reals). The tropical semiring is used for 
 minimal-cost computations. Suppose that the costs for establishing $\psi$ and $\phi$ are $s$ and $t$, respectively,
 and that you have already payed $s$ for $\psi$. Now you want to establish $\psi\ra\phi$ to infer $\phi$.
 If $s\geq t$ you have to pay nothing (implicitely you have already established $\phi$), otherwise you pay $t-s$.
 \item On the Łukasiewicz semiring $\Lukas= ([0,1]_\R, \max, \odot, 0, 1)$, where multiplication is given by $s \odot t = \max(s+t -1, 0)$,
 we have that $s\ra t$ is the Łukasiewicz implication $s\ra t:=\min(1-s+t,1)$.
 \item In the polynomial semirings $\PosBool[X]$ we can view each monomial as a set $Y\subseteq X$,
 multiplication of monomials corresponds to their union, and the empty set is 1. The natural order on monomials is defined by
 $Y\leq Z \Iff Y\supseteq Z$, in which case we say that $Z$ \emph{absorbs} $Y$.
 For monomials $Y,Z$ we thus have  $Y\ra Z= Z\setminus Y$. Each expression $p\in\PosBool[X]$ is an antichain of monomials,
 i.e. a collection of subsets $Y\subseteq X$ none of which is a subset of another one. Addition $p+q$ is defined by taking 
 all monomials in $p$ and $q$ and then deleting those that are absorbed by another one, keeping only the
 minimal subsets $Y\in p\cup q$.  The natural order on such expressions
 is that $p\leq q$ if every $Y\in p$ is a superset of some $Z\in q$; indeed it then follows that $p+q=q$ because
 every monomial $Y\in p\cup q$ is absorbed by some $Z\in q$.  It follows that $p\ra q$ is the collection
 of the $\subseteq$-minimal sets $U$ such that for each $Y\in p$, $U\supseteq Z\setminus Y$
 for some $Z\in  q$.   How difficult is this to compute?
 
 \erich{We should say something to the following:
 \item  For  $\PosBool[X,\bar X]$ with dual indeterminates ($\dots$)
 \item $\Sorb[X]$ and $\Sinf[X]$ (\dots)}
 \end{itemize}

 \begin{proposition}\label{DT} With this implication, the deduction theorem holds for every infinitary absorptive semiring. 
\end{proposition}
 
\begin{proof}  Assume $\Phi\models_\Semi \psi\ra\phi$. Then $\pi\ext{\Phi}\leq\sup\{r : r\cdot\pi\ext{\psi}\leq\pi\ext\phi\}$ for each $\Semi$-interpretation $\pi$,
so by monotonicity $\pi\ext{\Phi}\cdot\pi\ext{\psi}\leq\pi\ext\phi$. 
This proves that  $\Phi,\psi\models_\Semi \phi$.

Conversely assume  $\Phi,\psi\models_\Semi \phi$. For every  $\Semi$-interpretation $\pi$ we have 
$\pi\ext{\Phi}\cdot\pi\ext{\psi}\leq\pi\ext\phi$, which implies that $\pi\ext{\Phi}\leq\sup\{r: r\cdot\pi\ext{\psi}\leq\pi\ext\phi\}=\pi\ext{\psi\ra\phi}$.
This proves that  $\Phi\models_\Semi \psi\ra\phi$.
\end{proof}
 
\medskip\noindent\textbf{Negation. }
We can use the more general interpretation of $s\ra t$ defined above and put $\sim s:=(s\ra 0)$.
Accordingly we have a non-atomic negation operator on $FO^\ra$ with 
\[  \pi\ext{\sim\psi}:=\pi\ext{\psi\ra 0}=\sup\{ r: r\cdot\pi\ext{\psi}=0\}.\]

 \begin{itemize}
 \item On every semiring $\Semi$ without divisors of 0 we have 
 \[  \pi\ext{\sim\psi}=\begin{cases} 1 &\text{ if } \pi\ext{\psi}=0\\
 0 &\text{ otherwise.}\end{cases}\]
 \item On the  {\L}ukasiewicz semiring $\Lukas$ with the {\L}ukasiewicz implication $s\ra t:=\min(1-s+t,1)$,
 we have that $\pi\ext{\sim\psi}=1-\pi\ext{\psi}$.
\erich{\item $\PosBool[X,\bar X]$ \dots}
 \end{itemize}

\subsection{Double valuations}

Given that, under the standard approach, negation is not a compositional algebraic
operation, it makes sense to associate which each logical statement not 
one, but two semiring values, one for the statement itself, and one for its
negation. Given a semiring interpretation $\pi:\Lit_A(\tau)\ra \Semi$
we associate with each first-order statement $\psi\in\FO(\tau)$, which is not necessarily in negation normal form, a pair of values $\pi^*\ext\psi=(\psi^+,\psi^-)\in S\times S$
according to the following
inductive rules:
\begin{align*}
\pi^*\ext\alpha&:=(\pi(\alpha),\pi(\neg\alpha))\quad\text{if $\alpha$ is an atomic formula}\\ 
\pi^*\ext{\psi\lor\phi}&:=(\psi^+ + \phi^+, \psi^-\cdot\phi^-)\\
\pi^*\ext{\psi\land\phi}&:=(\psi^+ \cdot \phi^+, \psi^- + \phi^-)\\
\pi^*\ext{\E x\psi}&:=\Bigl(\sum\nolimits_{a\in A}\psi(a)^+,\prod\nolimits_{a\in A} \psi(a)^-\Bigr)\\ 
\pi^*\ext{\A x\psi}&:=\Bigl(\prod\nolimits_{a\in A}\psi(a)^+,\sum\nolimits_{a\in A} \psi(a)^-\Bigr)\\
\pi^*\ext{\neg \psi}&:=(\psi^-, \psi^+) 
\end{align*}

A simple induction shows that for every  $\psi\in\FO(\tau)$,
\[    \pi^*\ext\psi=(\pi\ext{\nnf(\psi)},\pi\ext{\nnf(\neg\psi)}).\]

For model-defining semiring interpretations into a positive semiring 
this is not particularly interesting
since we will always have that $\pi^*\ext{\psi}=(s,0)$ or 
 $\pi^*\ext{\psi}=(0,s)$ for some $s\in S\setminus\{0\}$.
However, for instance for model-compatible interpretations into
$\N[X,\nnX]$ a valuation gives insights into both proof trees and refutation trees,
or equivalently winning strategies for both players in the associated
model-checking game, depending on which literals are set to true.

\medskip This approach has so far not been explored any further, but we believe that
it deserves investigation.

\section{Further developments in semiring semantics}

The extension of semiring provenance from positive database queries to a semiring
semantics of full first-order logic, as proposed in \cite{GraedelTan17} 
has motivated further work in several directions, beyond the topics addressed above.

\subsection{Semiring semantics of fixed-point logic}
Semiring semantics has meanwhile be defined not only first-order logic, but for many other logics as well.
The most interesting challenges arise for fixed-point logics, such as LFP or the modal $\mu$-calculus. For one of the most simple fixed-point formalisms, namely the query language \emph{datalog}, a provenance analysis has already been provided in the original paper
\cite{GreenKarTan07}, and has later been extended in \cite{DeutchMilRoyTan14}. Due to the need of unbounded least fixed-point iterations 
in the evaluation of Datalog queries, the underlying semirings have to satisfy the additional property of being $\omega$-continuous,
which essentially means that ascending chains (with respect to the natural order) have a supremum that is compatible with the semiring operations. 
By Kleene's Fixed-Point Theorem, systems of polynomial equations then have least fixed-point solutions that can be computed by induction, reaching the fixed-point after at most $\omega$ stages. 
Most of the common application semirings are $\omega$-continuous, or can easily be extended to one that is so,  
but the general $\omega$-continuous provenance semiring over $X$  is no longer a semiring of polynomials but
the semiring of formal power series over $X$ (consisting of potentially infinite sums of monomials), denoted $\N^\infty[\![X]\!]$, with coefficients in $\Ninf\coloneqq\N\cup\{\infty\}$.
As above, provenance valuations $\pi\ext{\psi}\in\N^\infty[\![X]\!]$ give precise information about the possible
proof trees for a Datalog query. Even though 
the databases are assumed to be finite there
may be infinitely many proof trees, but each of them can use each atomic fact only a finite number of times, so the
proof trees are still finite. 

Our approach of extending polynomial semiring by dual indeterminates
readily extends to semirings of formal power series, which gives
us the semirings  $\N^\infty[\![X,\nnX]\!]$ of dual-indeterminate
power series \cite{GraedelTan20}. These are the general provenance semirings for semipositive Datalog (with negated input predicates) and, much more generally, also for $\posLFP$, the fragment of LFP that consists of formulae in negation normal form   
such that all its fixed-point operators are least fixed-points. This is a powerful fixed-point calculus,
that plays an important role in finite model theory and 
captures all polynomial-time computable properties of ordered finite structures \cite{Graedel+07}.

Nevertheless, for the general objective of a provenance analysis of fixed-point calculi,  the restriction to (positively used) 
least fixed points is not really satisfactory.   The transformation
from a fixed-point formula with arbitrary interleavings of least and greatest fixed points 
into one in $\posLFP$ is, contrary to transformations into negation normal form, 
not a simple syntactic translation. It goes through the Stage Comparison Theorem \cite{Moschovakis74,Graedel+07}
and can make a formula much longer and more complicated. Further, such transformations
are  not available for important fixed-point formalisms such as the modal $\mu$-calculus,
stratified Datalog, transitive closure logics, and even simple temporal languages
such as CTL.  
So the question arises what kind of semirings are adequate for a meaningful and informative
provenance analysis of unrestricted fixed-point logics, with arbitrary interleavings of least and greatest fixed points. This has been studied in \cite{DannertGraNaaTan21}.

Rather than just $\omega$-continuity one needs
semirings that are \emph{fully continuous} which means that every
chain has not only a supremum, but also an infimum,
and these are compatible with the algebraic operations of the
semiring.
For an informative provenance semantics, there is a second important condition that is 
connected with the symmetry, or duality, between least and greatest fixed points. Indeed, in the Boolean setting, a greatest 
fixed point of a monotone operator
is the complement of the least fixed of the dual operator (which is also monotone). It is this duality that
permits to push negations through to the atoms and work with formulae in negation normal form.
To have a similar kind of symmetry in the provenance setting, it is required that the semirings are
\emph{absorptive} so that multiplication is decreasing, i.e., $a \cdot b \le b$ for all $a,b$.
In particular, the powers of an element form a decreasing chain. 
It has been shown in \cite{DannertGraNaaTan21} that
absorptive and fully continuous semirings guarantee a well-defined and informative 
provenance semantics for arbitrary fixed-point formulae.
 
The most general absorptive and fully continuous semirings are the
semirings $\Sinf[X]$ of \emph{generalized absorptive polynomials}.  Informally such a polynomial is
a sum of monomials, with possibly infinite exponents, that are maximal with respect to absorption.
For instance a monomial $x^2y^\infty z$ occurring in a provenance value $\pi\ext{\psi}$ indicates
a model-checking proof that uses the atom labelled by $x$ twice,
the atom labelled by $y$ an infinite number of times, and the atom labelled by $z$ once. 
This monomial absorbs all those that have larger exponents for all variables, such as for instance $x^3y^\infty z^\infty u$,
but not, say, $x^\infty y^3$. Absorptive polynomials thus describe shortest model-checking proofs. Absorption has the further pleasant consequence that generalized formal power series collapse to generalized polynomials. Indeed we can view a polynomial or formal power series as an antichain of monomials (wrt. absorption order), and one can show \cite{GraedelTan20,DannertGraNaaTan21} that all these antichains are finite.

As for  semirings of polynomials, we can also in this case construct quotient semirings  
$\Sinf[X,\nnX]$ of dual indeterminate generalized polynomials to treat positive and negative atomic information appropriately, and these semirings
do indeed have universality properties that make them
the most general semirings for LFP \cite{DannertGraNaaTan21}.
An algorithmic analysis for computing least and greatest fixed point
in absorptive semirings has been given in \cite{Naaf21}.

\subsection{Strategy analysis of games}

Semiring provenance for logics is intimately related to semiring valuations of two-player games based on a correspondance that goes both ways.
On one side, evaluation problems for logical formulae can be
cast as the problem whether the verifying player has a winning strategy in an associated model checking game. On the other side
the winning region of a player (i.e., the set of positions from which
she has a winning strategy) can be defined by a formula in an appropriate logic (often in LFP). By computing valuations of game positions and strategies in an appropriate semiring, one can obtain
detailed information about the available strategies of a player, far beyond the fact, that a player will win or lose.

The mathematical basis of a strategy analysis via semiring valuations are \emph{Sum-of-Strategies Theorems}. The ingredients of such theorems are:
\begin{itemize}
    \item A valuation of game positions and/or moves by elements of a semiring $\Semi$. In an acyclic game, which admits only finite plays,  this is done by a simple backwards induction; in more complicated games such as B\"uchi or parity games, this can be defined by a valuation of a formula which states that there is a winning strategy from the given position.
    \item An appropriate class of strategies for the game.
    \item A valuation of these strategies by elements of the semiring $\Semi$; normally this is the product of the valuations of the moves or positions that occur in the strategy.
\end{itemize}

The Sum-of-Strategies Theorem then says that the semiring valuation 
of any position in the game coincides with the sum of the valuations
of the available strategies from that position. 

The simplest instance of this concerns the evaluation of a first-order sentence on a finite structure or more generally, on a
semiring interpretation. A winning strategy in the associated model checking game is precisely the same as a proof tree, and the Sum-of-Strategies Theorem for first-order model checking games is
just a different way to state the Sum-of-Proof-Trees Theorem
for FO (Theorem~\ref{thm:prooftrees} in Sect.~\ref{sec:prooftrees}).
More complicated instances concern acyclic reachability games
\cite{GraedelTan20} and the model checking games for LFP \cite{DannertGraNaaTan21}. 

A specific case study for evaluating the power of this approach
has been done in \cite{GraedelLucNaa21} for Büchi games. These are games with a winning condition requiring that some good position is seen infinitely often during the play. Büchi games have a number of practical applications, but they are also of interest because they are among the simplest games where
any formula win$(v)$ saying that $v$ is a winning position requires a genuine nesting of a least fixed point inside a greatest fixed point. The approriate class of strategies in this case are
the \emph{absorption-dominant} strategies (strategies that win with minimal effort), and the Sum-of-Strategies Theorem 
states that for any position $v$ in a B\"uchi game, the
valuation of the LFP-formula $\textrm{win}(v)$ in an absorptive, fully-continuous semiring
coincides with the sum of the valuations of all absorption-dominant winning strategies from $v$.
From such a valuation on can derive not only whether a player wins from $v$, but also the number and  shapes of all absorption-dominant
(as also all positional) winning strategies. Further one can determine whether a player still wins if certain moves are forbidden,
or must be used only finitely often. Finally, such valuations also
can be used to repair a game, i.e. to find minimal changes of a game that
cannot be won into one that admits a winning strategy, by techniques
that are similar to the ones in Sect.~\ref{sec:repairs}.

\subsection{The model theory of semiring semantics}

The development of semiring semantics for various logics, and specifically for full first-order logic, raises the question to what extent classical techniques and results of logic extend to semiring semantics, and how this depends on the algebraic properties of the underlying semiring. In a general research programme that
explores such questions, the following topics have been investigated so far.

\subsubsection{Elementary equivalence versus isomorphism} 

It is a rather obvious logical fact that every finite 
structure (with a finite vocabulary) can be axiomatised, 
up to isomorphism, by a first-order sentence. 
In particular, two finite $\tau$-structures $\AA$ and $\BB$ are isomorphic if,
and only if, they are elementarily equivalent, in short $\AA \equiv \BB$, 
which means that they cannot be 
distinguished by any first-order sentence. Is this also the case for semiring interpretations?
Notice that standard notions such as isomorphism and elementary equivalence generalise in
a natural way from $\tau$-structures to semiring interpretations, which raises,
for any given semiring $\Semi$, the following questions.

\begin{enumerate}
\item Are elementary equivalent finite $\Semi$-interpretations always isomorphic?\label{item:question1}
\item Is every finite $\Semi$-interpretation $\pi_A$ first-order axiomatisable, i.e. is there is a set of axioms 
$\Phi_A \subseteq \FO$ such that whenever
$\pi_B \ext{\phi} = \pi_A \ext{\phi}$ for all $\phi \in \Phi_A$, then $\pi_B \cong \pi_A$?\label{item:question2}
\item Does every finite $\Semi$-interpretation admit an axiomatisation by a \emph{finite} set of axioms?\label{item:question3}
\item Can every finite $\Semi$-interpretation be axiomatised by a single first-order sentence?\label{item:question4}
\end{enumerate}

Clearly, the first two questions are equivalent, and a
positive answer to the third question implies also positive ones to the first two.
The converse is not necessarily true, because a first-order axiomatisation of a finite
semiring interpretation might require an infinite collection of sentences,
and, contrary to the Boolean case,
it is a priori also not clear that an axiomatisation by a finite set of sentences implies
an axiomatisation by a single sentence, because from the
value of a conjunction we cannot necessary infer the values of its components.

It has been shown in \cite{GraedelMrk21} that the answers to these questions strongly depend on the chosen semiring.
There are in fact rather simple semirings, such as min-max semirings with at least three elements,
for which one can construct examples of non-isomorphic interpretations which
are, however, elementarily equivalent. 
Since the standard method for establishing elementary equivalence is not generally available
in semiring semantics (see Sect.~\ref{sect:EF}), new methods
based on separating sets of semiring homomorphisms had to be developed
for this purpose.
Elementarily equivalent but non-isomorphic semiring interpretations also exist
for provenance semirings, such as $\Sorb(X)$, $\Bool[X]$ and $\Why(X)$.
On the other side, there are  semirings, such as the Viterbi semiring $\Vit$,
the tropical semiring $\Trop$, the natural semiring $\N$ and the universal polynomial
semiring $\N[X]$, for which any finite interpretation is first-order axiomatisable,
so that  elementary equivalence does indeed imply
isomorphism. At least for $\Vit$ and $\Trop$, finite axiomatisations are always possible,
but not axiomatisations by a single sentence, so there exist semirings where the answers
to \hyperref[item:question3]{questions (3) and (4)} are different.

\subsubsection{0-1 laws}
  
The classical 0-1 law for first-order logic, due to
Glebskii et al. \cite{GlebskiiKogLioTal69} and Fagin
\cite{Fagin76}, says that the probabilities that a relational 
first-order sentence is true in a random finite  
structure converge exponentially fast to either 0 or 1, as the size of the structures grows to infinity. Informally speaking, 
on random finite structures, every first-order sentence is either almost surely false or almost surely true.
Random semiring interpretations, induced by a probability distribution on the non-zero elements of a semiring, generalise random structures, and the question arises whether also the 0-1 law generalise to semiring semantics.
This has been studied in  \cite{GraedelHelNaaWil22} with the following results. 

On many different semirings $\Ss$ there indeed is a 0-1 law, saying that with probabilities converging to 1 exponentially fast,  the valuation $\pi\ext{\psi}$ of a first-order sentence $\psi$ almost surely concentrates on one specific value $s\in \Ss$.
This induced a partition  of $\FO(\tau)$ into classes 
 $(\Phi_s)_{s\in S}$ such that sentences in $\Phi_s$ evaluate
almost surely to $s$. 
On finite lattice semirings, this partition collapses to just three classes $\Phi_0$, $\Phi_1$, and $\Phi_\epsilon$, of sentences that respectively, almost surely 
evaluate to 0, 1, and to the smallest value $\epsilon\neq 0$.
For all other values $s\in S$ we have that
$\Phi_s=\emptyset$. The problem of computing the almost sure valuation of a first-order sentence on
finite lattice semirings is {\sc Pspace}-complete.

The methods to prove such results combine on the one hand techniques that are adapted from traditional studies of logic on random structures, such as extension properties of atomic types, 
and on the other side specific ideas of semiring semantics, such as a specific variant of provenance tracking polynomials.

A  semiring where the analysis is somewhat different is the \emph{natural semiring} $(\N,+,\cdot,0,1)$.
The 0-1 law still holds for the natural semiring, but the proof relies on more general $\infty$-expressions instead of polynomials and there are rather trivial constructions showing that every number $j\in\N$ appears as a possible almost sure valuation.

\subsubsection{Locality} 
Locality is a fundamental property of first-order logic 
and an important limitation of its expressive power.  Informally, this
means that the truth of a first-order formula $\psi(\bar x)$ in a given structure only depends on a 
neighbourhood of bounded radius around $\bar x$, and on the existence of a bounded number of local substructures.
Consequently, first-order logic cannot express global properties such as connectivity or acyclicity of graphs.
Two fundamental theorems that make this precise are
\emph{Hanf's locality theorem} and \emph{Gaifman's normal form theorem}.
In a nutshell, Hanf's theorem gives a criterion for the $m$-equivalence 
(i.e. indistinguishability by sentences of quantifier rank up to $m$) of two structures  
based on the number of  local substructures of any given isomorphism type, while
Gaifman's theorem
states that every first-order formula is equivalent to a Boolean combination of local formulae and basic local sentences;
this has many model-theoretic and algorithmic consequences. 
The question whether such locality theorems also hold in semiring semantics has been studied in \cite{BiziereGraNaa23}.
It is shown that
Hanf's theorem generalises to all semirings where both operations are idempotent, but fails for many other semirings.
For formulae with free variables, Gaifman's theorem does not generalise beyond the Boolean semiring, and
also for sentences, it fails in some important semirings such as the natural semiring and the tropical semiring.
The main result, however, is a constructive proof of the existence of Gaifman normal forms for min-max and lattice semirings.
In fact, this proof also implies a stronger version of Gaifman's classical theorem in Boolean semantics, saying that
every sentence has a Gaifman normal form which does not add negations.

\subsubsection{Ehrenfeucht--Fraïssé games}\label{sect:EF} 
To prove elementary equivalence (and equivalence up to
a fixed quantifier rank) of relational structures a standard method (in particular in finite model theory)
is provided by Ehrenfeucht--Fraïssé games or, equivalently, back-and-forth systems of local isomorphisms.
But while Ehrenfeucht--Fraïssé games are sound and complete for logical equivalences in classical semantics, and thus on the Boolean semiring,
this is in general not the case for other semirings. A detailed analysis of the soundness and completeness of model comparison games
on specific semirings, not just for classical Ehrenfeucht--Fraïssé games but also 
for other variants based on bijections or counting, is provided in \cite{BrinkeGraMrk24}. 
It turns out that $m$-move Ehrenfeucht--Fraïssé games are sound
(but in general not complete) for $m$-equivalence on fully idempotent semirings, whereas $m$-move bijection games are sound on all semirings.
Ehrenfeucht--Fraïssé games without a fixed restriction on the number of moves are sound for elementary equivalence on a
number of further important semirings, but completeness only holds in rare cases.
Based on the results in \cite{GraedelMrk21} that there exist certain rather simple semiring interpretations that are locally very different and 
can be separated even in a one-move game, but which can be proved to be elementarily equivalent via separating sets of homomorphisms,
a new kind of games, called  \emph{homomorphism games} has been developed in \cite{BrinkeGraMrk24},
which provide a sound and complete method for logical equivalences on finite lattice semirings.

\section{Acknowledgements}

Our collaboration on the topics of this paper started in Fall 2016 as we were both participating in the ``Logical Structures in Computation'' program at the Simons Institute for the Theory of Computing in Berkeley. We are very grateful to the Institute for support and for the perfect collaborative atmosphere that it fosters. We would like to acknowledge very useful discussions at the Institute with Andreas Blass, Mikolaj Bojanczyk, Thomas Colcombet, Anuj Dawar, Kousha Etessami, Diego Figueira, Phokion Kolaitis, Ugo Montanari, Jaroslav Nesetril, Daniela Petrisan and Miguel Romero.

Val Tannen is very grateful to his collaborators in the development over several years of semiring provenance for databases: (in chronological-alphabetical order) T.J. Green, Grigoris Karvounarakis, Zack Ives, Nate Foster, Yael Amsterdamer, Daniel Deutch, Tova Milo, Susan Davidson, Julia Stoyanovich, Sudeepa Roy, Yuval Moskovitch, Abdu Alawini, Jane Xu, and Waley Zhang. He was partially supported by NSF grants 1302212 and 1547360 and by NIH grant U01EB02095401.


\bibliography{provenance}

\end{document}